\documentclass[preprint,aps,12pt,showpacs,nofootinbib,tightenlines,amsmath,amssymb]{revtex4}
\usepackage{amsmath}
\usepackage{graphicx}
\usepackage{CJK,upgreek,fancyhdr}
\usepackage{color}

\newcommand{\btd}{\bigtriangledown}
\newcommand{\be}{\begin{eqnarray}}
\newcommand{\ee}{\end{eqnarray}}

\newcommand{\nn}{\nonumber}

\newcommand{\diag}{\mathop{\rm diag}}

\newcommand{\Gm}{\mathsf \Gamma}
\newcommand{\dg}{\digamma}

\newcommand{\Wm}{\mathsf W}

\newcommand{\chih}{\hat{\chi}}

\newcommand{\ha}{\hat{a}}
\newcommand{\hb}{\hat{b}}
\newcommand{\ch}{\hat{c}}
\newcommand{\hd}{\hat{d}}
\newcommand{\he}{\hat{e}}
\newcommand{\hf}{\hat{f}}

\newcommand{\ta}{\tilde{a}}

\newcommand{\hx}{\hat{x}}
\newcommand{\hm}{\hat{\mu}}
\newcommand{\hn}{\hat{\nu}}
\newcommand{\hr}{\hat{\rho}}

\newcommand{\hs}{\hat{\sigma}}
\newcommand{\tc}{\tilde{c}}
\newcommand{\tC}{\tilde{C}}
\newcommand{\hc}{\hat{c}}
\newcommand{\hC}{\hat{C}}
\newcommand{\tm}{\tilde{\mu}}

\newcommand{\tx}{\tilde{x}}

\begin{document}
\begin{CJK*}{GB}{gbsn}
\def\intdk{\int\frac{d^4k}{(2\pi)^4}}
\def\sla{\hspace{-0.17cm}\slash}
\hfill


\title{Maximal symmetry and mass generation of Dirac fermions and gravitational gauge field theory in six-dimensional spacetime}

\author{Yue-Liang Wu (ÎâÔÀÁ¼)}\email{ylwu@itp.ac.cn; ylwu@ucas.ac.cn}
\affiliation{CAS Key Laboratory of Theoretical Physics \\
Institute of Theoretical Physics, Chinese Academy of Sciences, Beijing, 100190, China\\
International Centre for Theoretical Physics Asia-Pacific (ICTP-AP) \\
University of Chinese Academy of Sciences (UCAS), Beijing 100049, China }


\begin{abstract}
The relativistic Dirac equation in four-dimensional spacetime reveals a coherent relation between the dimensions of spacetime and the degrees of freedom of fermionic spinors. A massless Dirac fermion generates new symmetries corresponding to chirality spin and charge spin as well as conformal scaling transformations. With the introduction of intrinsic W-parity,  a massless Dirac fermion can be treated as a Majorana-type or Weyl-type spinor in a six-dimensional spacetime that reflects the intrinsic quantum numbers of chirality spin. A generalized Dirac equation is obtained in the six-dimensional spacetime with a maximal symmetry. Based on the framework of gravitational quantum field theory proposed in Ref.~\cite{YLWU} with the postulate of gauge invariance and coordinate independence, we arrive at a maximally symmetric gravitational gauge field theory for the massless Dirac fermion in  six-dimensional spacetime. Such a theory is governed by the local spin gauge symmetry SP(1,5) and the global Poincar\'e symmetry P(1,5)= SO(1,5)$\ltimes P^{1,5} $ as well as the charge spin gauge symmetry SU(2). The theory leads to the prediction of doubly electrically charged bosons. A scalar field and conformal scaling gauge field are introduced to maintain both global and local conformal scaling symmetries. A generalized gravitational Dirac equation for the massless Dirac fermion is derived in the six-dimensional spacetime. The equations of motion for gauge fields are obtained with conserved currents in the presence of gravitational effects. The dynamics of the gauge-type gravifield as a Goldstone-like boson is shown to be governed by a conserved energy-momentum tensor, and its symmetric part provides a generalized Einstein equation of gravity.  An alternative geometrical symmetry breaking mechanism for the mass generation of Dirac fermions is demonstrated. 
\end{abstract}

\maketitle

\section{introduction}
 
After building the theory of general relativity~\cite{GR}, Einstein predicted the existence of gravitational waves~\cite{GW}. The direct observation of gravitational waves at LIGO~\cite{LIGO} and the discovery of the Higgs boson at the LHC \cite{higgs2012a,higgs2012c} motivate us to further explore more fundamental issues, such as the structure of matter, the dynamics of spacetime, the origin of the universe  and the mass generation of quantum fields. These are the most challenging problems in the basic sciences and all concern a deep understanding of the nature of gravity. 

The gravitational force is currently characterized by the theory of general relativity (GR), which is applied successfully to describe the macroscopic world. GR was formulated by Einstein based on the postulate: {\it the laws of physics must be of such a nature that they apply to systems of reference in any kind of motion}. More explicitly, the physical laws of nature are to be expressed by equations which hold good for all systems of coordinates, 
\be
R_{\mu\nu}-\frac{1}{2} g_{\mu\nu} R =  \frac{8\pi G}{c^4} T_{\mu\nu} \, , \quad G = \frac{\hbar c^5}{M_P^2}\, , \quad (\mu = 0,1,2,3)
\ee
where $G$ is Newton's gravitational constant and $c$ the speed of light in vacuum, and $\hbar$ and $M_P$ the Planck constant and Planck mass, respectively. The left-hand side shows the geometric property of spacetime with $R_{\mu\nu}$ the Ricci tensor and $R = g^{\mu\nu} R_{\mu\nu}$ the curvature tensor, which is determined by the metric tensor $g_{\mu\nu}$, and the right-hand side reflects the property of matter with $T_{\mu\nu}$ the energy-momentum tensor of matter.

The establishment of GR and its experimental tests have led to a breakthrough in the understanding of the structure of spacetime and the correlation between the geometric property and matter distribution. It has been shown that the gravitational force is characterized as the dynamic Riemannian geometry of curved space-time. Namely, the physical laws are invariant under the general linear transformation of local GL(4,R) symmetry, which indicates that space and time can no longer be well defined in such a way that the differences of the spatial coordinates or time coordinate can be directly measured by the standard ways proposed in special relativity. Furthermore, energy-momentum conservation cannot be well defined, as the GL(4,R) group contains no symmetry of the translational group $P^{1,3}$. Manifestly, it is in contrast to other three basic forces, the electromagnetic, weak and strong forces~\cite{QED1, QED2, QED3, QED4, EW1, EW2, EW3, QCD1, QCD2}, which are all described by gauge field interactions within the framework of quantum field theory (QFT). QFT has been established based on the globally flat Minkowski spacetime of special relativity and possesses a global symmetry of inhomogeneous Lorentz groups in four-dimensional spacetime, which contains both the Lorentz group SO(1,3) and translational group $P^{1,3}$, i.e., Poincar\'{e} group P(1,3) = SO(1,3)$\ltimes P^{1,3}$. 

Such an odd dichotomy causes difficulties for the quantization of gravitational force and the unification of gravitational force with the other three basic forces. This should not be surprising, as GR was formulated by Einstein in 1915 as a direct extension to special relativity. At that time, quantum mechanics was not yet established and the equation of motion of the electron was unclear, not to mention that the weak and strong forces as well as the basic building blocks of nature and the framework of QFT were all unknown. Quantum mechanics was originally established in the mid-1920s as a non-relativistic quantum mechanics and obtained by quantizing the equations of classical mechanics by replacing dynamical variables with operators. Its mathematical formulation is applied in the context of Galilean relativity. 

A relativistic formulation was a natural extension of non-relativistic quantum mechanics. The key progress made by Dirac was the relativistic quantum Dirac equation~\cite{DE},
\be
\left(\mathsf{E} - c \boldsymbol{\alpha}\cdot \mathsf{\bf{p}} - \beta mc^2 \right) \psi = 0 \, ; \quad \mbox{or}\quad  \left( \mathsf{E} + c \boldsymbol{\alpha}\cdot \mathsf{\bf{p}} + \beta mc^2 \right) \psi =0 
\ee
with $\boldsymbol{\alpha} =(\alpha_1, \alpha_2, \alpha_3 )$ and $\beta$ the $4\times 4$ Hermitian matrices satisfying the conditions $\alpha _{i}\beta =-\beta \alpha _{i}$, $\alpha _{i}\alpha _{j}=-\alpha _{j}\alpha _{i}$ ( $i\neq j $) and $\alpha _{i}^{2}=\beta ^{2}= 1 $. The matrices multiplying $\psi$ shows that it is a field with a complex four-component entity. The Dirac equation is relativistically invariant. More explicitly, a Poincar\'{e} covariant formulation of the Dirac equation can be expressed as
\be \label{DiracEQ}
& & \left( \gamma^{\mu} i \partial_{\mu}  - m \right) \psi = 0\, , \qquad \quad \mu = 0,1,2,3, \, \nn \\
& & \psi^T =(\psi_1\, , \psi_2\, ,\psi_3\, , \psi_4) \, , \qquad \gamma^0 = \beta\, , \quad \gamma^i = \beta\alpha^i\, , 
\ee
with the 4-dimensional coordinate derivative $\partial_{\mu} = \partial/\partial x^{\mu}$ and the 4$\times$4 $\gamma$-matrix $\gamma_{\mu}$, which satisfies anticommutation relations 
\be
\{ \gamma_{\mu}\, \,  \gamma_{\nu} \} = \eta_{\mu\nu}\, , \qquad \eta_{\mu\nu} = \diag.(1, -1, -1, -1)
\ee

The Dirac equation is a crucial step for the unity of  quantum mechanics and special relativity, which led to the successful development of relativistic quantum mechanics and QFT. The Dirac equation reveals an interesting correlation between the dimensions of spacetime and the degrees of freedom or quantum numbers of the basic building blocks, such as electrons and quarks. Specifically, the four-dimensional spacetime of coordinates correlates with the four-component entity of the Dirac spinor.  Such a four-component entity field reflects the spin-$\frac{1}{2}$ property and negative energy solution of the Dirac field $\psi(x)$, which provided a pure theoretical prediction of the existence of antiparticles. 

Inspired by the relativistic Dirac equation, in this paper we will investigate  the intrinsic properties of a massless Dirac spinor and show an additional correlation between the chirality spin quantum numbers and extra dimensions. We shall begin with a demonstration to show the maximal symmetry for a massless Dirac spinor and how to derive a generalized Dirac equation in a six-dimensional spacetime. We then follow the framework of gravitational quantum field theory proposed in Ref.~\cite{YLWU} with the postulate of gauge invariance and coordinate independence, which is an alternative to Einstein's postulate on general covariance of coordinate transformation for the theory of general relativity~\cite{GR}, to find a unified description of quantum mechanics and gravitational interaction for the massless Dirac spinor in the six-dimensional spacetime.  

Our paper is organized as follows. In Section 2, by considering a massless Dirac spinor, we extend the Dirac equation in four-dimensional spacetime to an extended Dirac equation in a six-dimensional spacetime, so that the Lorentz symmetry group SO(1,3) in the vector representation of coordinates and the spinor spin symmetry group SP(1,3)$\cong$ SO(1,3) in the spinor representation of the Dirac field are generalized to the Lorentz symmetry group SO(1,5) and the spinor spin symmetry group SP(1,5), respectively. Under the ordinary parity and time-reversal operations, the fifth and sixth dimensions transform as spatial and time-like dimensions, respectively. An intrinsic W-parity is introduced to characterize the two extra dimensions. With the properties of charge-conjugation and W-parity,  we demonstrate in Section 3 the existence of a maximal internal gauge symmetry SU(2) for the massless Dirac spinor. An action with the maximal symmetry is built in the globally flat six-dimensional Minkowski spacetime. We then derive a generalized relativistic quantum equation for the massless Dirac field and the gauge field as well as the singlet scalar field introduced to maintain a conformal scaling symmetry. Doubly electrically charged bosons are predicted in the theory. In Section 4,  when taking the spinor spin group symmetry SP(1,5) and the global conformal scaling symmetry of the Dirac field as internal gauge symmetries, we are led to a biframe spacetime structure $T_M\times G_M$ associated with its dual spacetime $T_M^{\ast}\times G_M^{\ast}$ over the Minkowski spacetime $M_A$. A bicovariant vector gravifield $\chih_{\ha}^{\;\;\hm}(x)$ (a dual gravifield $\chi^{\;\; \ha}_{\hm}(x)$) defined on $T_M\times G_M$ (a dual biframe spacetime $T_M^{\ast}\times G_M^{\ast}$) is introduced to characterize the gravitational interaction. Based on the nature of globally and locally flat vector spacetimes,  it allows for the canonical identification of vectors in the tangent Minkowski spacetime $T_M$ at points with vectors in the Minkowski spacetime $M_A$ itself, and also the canonical identification of vectors at a point with its dual vectors at the same point.  The total spacetime is viewed as a {\it gravifield fiber bundle} $\bf{E}$ with the identified locally flat {\it gravifield spacetime} ${\bf G} \equiv G_M \cong G_M^{\ast} $ as the fiber and the identified globally flat {\it vacuum spacetime} ${\bf V} \equiv T_M \cong T_M^{\ast} \cong  M_A $ as the base spacetime. With the principle of gauge invariance and coordinate independence proposed recently in Ref.~\cite{YLWU} for a quantum field theory of gravity, we arrive at a gravitational gauge field theory for a massless Dirac spinor with maximal symmetry in the locally flat gravifield spacetime ${\bf G}$. In Section 5,  we present an alternative formalism by projecting the action into the globally flat Minkowski  spacetime ${\bf V}$, which enables us to derive equations of motion for all fields and conservation laws for all symmetries. The equation of motion for the gauge-type gravifield is in general connected with a nonconserved current. When turning to a hidden gauge formalism, the dynamics of the gravifield is found to be characterized by a total energy-momentum tensor. The conservation of total energy-momentum tensor leads to an interesting relation between the field strengths and the spacetime gauge field. The symmetric part of the equation of the gravifield tensor gives a generalized Einstein equation of gravity in the six-dimensional spacetime. Our conclusions and remarks are given in Section 6.

\section{Chirality spin \& W-Parity of Massless Dirac spinor and Extra Dimensions with Maximal Lorentz \& Spin Symmetry}

To show the explicit symmetries of a theory, it is useful to write down a corresponding action for such a theory. The action to yield the Dirac equation (\ref{DiracEQ}) can simply be written as 
\be \label{Dirac4D}
S_{D}^{4d} = \int d^4x\,  \frac{1}{2} \left( \bar{\psi}(x) \gamma^{\mu} i \partial_{\mu} \psi(x) + H.c. \right)  - m \bar{\psi}(x) \psi(x)
\ee
with $\bar{\psi}(x) = \psi^{\dagger}(x) \gamma^0$ as antispinor field.  The action is invariant under the global Poincar\'{e}  group P(1,3) = SO(1,3)$\ltimes P^{1,3}$. The global Lorentz group transformation is given by
\be
& & x^{\mu} \to x^{'\mu} = L^{\mu}_{\; \;\; \nu}\; x^{\nu}, \qquad  \psi(x) \to \psi'(x') = S(L) \psi(x), \nonumber \\
& &  L^{\mu}_{\; \;\; \nu} L^{\rho}_{\; \;\; \sigma} \eta_{\mu\rho} = \eta_{\nu\sigma}\; , \qquad \quad  L^{\mu}_{\; \;\; \nu} \in SO(1,3)\; 
\ee
with the group element
\be
& &  S(L) = e^{i\alpha_{\mu\nu} \Sigma^{\mu\nu}/2}  \in SP(1,3), \quad  \Sigma^{\mu\nu} = \frac{i}{4}[\gamma^{\mu}, \gamma^{\nu}] \, \nonumber \\
& &  S(L) \gamma^{\mu} S^{-1}(L) = L^{\mu}_{\;\;\; \nu}\; \gamma^{\nu}\; , 
\ee  
where $\Sigma^{\mu\nu}$ are the generators of the spin group SP(1,3) in the spinor representation. From the isomorphism property of the group,  the spin group SP(1,3) is isomorphic to the Lorentz group SO(1,3), i.e., SP(1,3) $\cong$ SO(1,3). For the translational group $P^{1,3}$, it is invariant under the parallel translation of coordinates
\begin{eqnarray}
 x^{\mu} \to x^{'\mu} = x^{\mu} + a^{\mu}
 \end{eqnarray}
with $a^{\mu}$ the constant vector. 

It indicates that the external rotational invariance of spacetime coordinates in the vector representation is coherent to the internal spin invariance of the Dirac field in the spinor representation. Namely, the external symmetry SO(1,3) of the four-dimensional spacetime coordinates must coincide with the internal symmetry SP(1,3) of the four-component entity Dirac field. As a consequence, it results in the conservation of total angular momentum. Specifically, the internal spin symmetry SP(1,3) incorporates a {\it boost spin} symmetry SU$^{\ast}$(2) and a {\it helicity spin} symmetry SU(2), i.e., SP(1,3)$\cong$SO(1,3)$\cong$SU$^{\ast}$(2)$\times$SU(2).

The mass of a Dirac spinor is supposed to originate from spontaneous symmetry breaking. Let us consider a massless Dirac spinor $m=0$. As a consequence, the above action generates two new symmetries.  

One is the global {\it conformal scaling symmetry}. Namely, the action is invariant under the global conformal scaling transformation for the coordinates and Dirac field,
\begin{eqnarray}
x^{\mu} \to x^{'\mu} = \lambda^{-1}\; x^{\mu},  \quad \psi(x) \to \psi'(x') = \lambda^{3/2}\; \psi(x), 
\end{eqnarray}
with $\lambda$ the constant scaling factor. 

The other is the so-called {\it chiral symmetry}. It can be shown that the action is invariant under the global chiral transformation 
\be
\psi(x) \to \psi'(x) = e^{i \alpha_5 \gamma_5} \psi(x) 
\ee 
with 
\be
\gamma_5 \gamma_{\mu} = - \gamma_{\mu} \gamma_5, \quad \gamma_5 = i \gamma^0\gamma^1\gamma^2\gamma^3 =
-\frac{i}{4!} \epsilon_{\mu\nu\sigma\rho}\gamma^{\mu}\gamma^{\nu}\gamma^{\sigma}\gamma^{\rho}\, .
\ee
It characterizes an intrinsic property of the Dirac spinor, which may be called a {\it chirality spin}.

Inspired by the derivation of the Dirac equation, it is natural to postulate that such a {\it chirality spin} invariance of massless Dirac spinors in the spinor representation reflects a rotational invariance of extra spacetime dimensions.  It is not difficult to check that once the Dirac spinor becomes massless $m=0$, the action Eq.~(\ref{Dirac4D}) in four-dimensional spacetime can be extended to an action in a six-dimensional spacetime,
\be \label{Dirac6D}
S_{D}^{6d} = \int d^6 x\,  \frac{1}{2} \left( \bar{\psi}(\hat{x}) \Gamma^{\hat{\mu}} i \partial_{\hat{\mu}} \psi(\hat{x}) + H.c. \right) \, , \quad \hat{\mu} = (\mu, 5, 6)\, ,
\ee
where $ \psi(\hat{x})$ remains a four-component entity Dirac spinor but as a field of six-dimensional spacetime coordinates $\hx=\{x^{\hat{\mu}} \} $. We have used the following notations,
\be 
& & x^{\hm} = (x^{\mu}, x^5, x^6)\, , \qquad  \Gamma^{\hm} = (\gamma^{\mu}, \Gamma^5, \Gamma^6)\, , \quad \Gamma^5 = i \gamma^5\, , \; \; \Gamma^6 = I_4 \, , 
\ee
with $I_4$ denoting the 4$\times$4 unit matrix. The equation of motion for such a massless Dirac spinor reads
\be
\Gamma^{\hat{\mu}} i \partial_{\hat{\mu}} \psi(\hat{x}) = 0\, , \qquad \eta^{\hm\hn}\partial_{\hat{\mu}} \partial_{\hat{\nu}}\psi(\hat{x}) = 0\, , 
\ee
with the constant metric matrix
\be
\eta^{\hm\hn} = \eta_{\hm\hn} =\diag.(1, -1,-1,-1,-1,-1)\, .
\ee

The signature of  $\eta_{\hm\hn}$ is $-4$, which indicates that the additional two dimensions are spatial ones. This can explicitly be checked from the invariance of chiral transformation of the Dirac spinor. The coordinates of the fifth and sixth dimensions transform correspondingly as a rotation, i.e.,  
\be
\psi(x) \to \psi'(x) = e^{i \alpha \gamma_5} \psi(x) \, , \quad 
 \binom{x_5}{x_6} \to  \binom{x'_5}{x'_6} =  \begin{pmatrix} 
 \cos\alpha & \sin\alpha \\
 -\sin\alpha & \cos \alpha 
 \end{pmatrix}
  \binom{x_5}{x_6} \, .
\ee

It can be shown that the action, Eq.~(\ref{Dirac6D}), becomes invariant under the Lorentz group SO(1,5)
\be
& & x^{\hm} \to x^{' \hm} = L^{\hm}_{\; \;\; \hn}\; x^{\hn}, \quad  \psi(\hx) \to \psi'(\hx') = S(L) \psi(\hx)\, , \nonumber \\
& &  L^{\hm}_{\; \;\; \hn} L^{\hr}_{\; \;\; \hs}\, \eta_{\hm\hr} = \eta_{\hn\hs}\; ,  \quad L^{\hm}_{\; \;\; \hn} \in SO(1,5) \, ,
\ee
where $S(L)$ is the spin group element in the spinor representation
\be
& &  S(L) = e^{i\alpha_{\hm\hn} \Sigma^{\hm\hn}/2}  \in SP(1,5), \quad   S(L) \Gamma^{\hm} S^{-1}(L) = L^{\hm}_{\;\;\; \hn}\; \Gamma^{\hn}  \, \nonumber \\ 
& & \Sigma^{\hm\hn} = ( \Sigma^{\mu\nu}, \, \Sigma^{\mu 5}, \, \Sigma^{\mu 6},\, \Sigma^{5 6} ) \, ,  \quad  \Sigma^{\mu\nu} = \frac{i}{4}[\gamma^{\mu}, \gamma^{\nu}] \, , \nonumber \\
& & \Sigma^{\mu 5} = -\Sigma^{5\mu} = -\frac{1}{2}\gamma^{\mu}\gamma^5, \quad \Sigma^{\mu 6} = -\Sigma^{6\mu} = \frac{1}{2} i \gamma^{\mu}, \quad \Sigma^{56} = -\Sigma^{65} = -\frac{1}{2} \gamma^5  \; , 
\ee  
where the fifteen $4\times 4$ matrices $\Sigma^{\hm\hn}$ are the generators of spin group SP(1,5) in the spinor representation. The transformation under the internal spin group SP(1,5) has to coincide with that under the external Lorentz group SO(1,5) in order to preserve maximal symmetry.

From group isomorphism, SP(1,5)$\cong$SO(1,5)$\cong$SU$^{\ast}$(4), the spin group SP(1,5) provides a maximal unitary symmetry for the four-component entity complex Dirac field. To further reveal intrinsic properties of the above action, we shall demonstrate how the extra two dimensions transform under ordinary parity-inversion ($\mathcal{P}$), time-reversal ($\mathcal{T}$) and charge-conjugation ($\mathcal{C}$). To make the action invariant and nontrivial under the discrete symmetries $\mathcal{P}$, $\mathcal{T}$ and $\mathcal{C}$ in the six-dimensional spacetime, the Dirac spinor as a field of six-dimensional spacetime coordinates should transform as follows:
\be
& & \mathcal{P} \psi(\hx ) \mathcal{P}^{-1} = P \psi( x^0,-x^k, -x^5, x^6 )\, , \quad k=1,2,3 \, , \nonumber \\
& &  P^{-1} \Gamma^{\hm} P = \Gamma^{\hm \dagger}, \quad P=\gamma^0 \, , 
\ee
for parity-inversion, 
\be
& & \mathcal{T} \psi(\hx ) \mathcal{T}^{-1} = T \psi( -x^0, x^k, -x^6 )\, , \quad k=1,2,3,5 \, , \nonumber \\
& &  T^{-1} \Gamma^{\hm} T = \Gamma^{\hm\, T}, \quad T= i \gamma^1\gamma^3\, , 
\ee
for time-reversal, and 
\be
& & \mathcal{C} \psi(\hx ) \mathcal{C}^{-1} = C \bar{\psi}^{T} ( x^{\mu},  -x^5, -x^6 )\, , \quad \mu=0,1,2,3 \, , \nonumber \\
& & C^{-1} \Gamma^{\mu} C = - \Gamma^{\mu\, T}, \quad  C^{-1} \Gamma^{k } C = \Gamma^{k \, T}, \quad k = 5,6,  \quad C= i \gamma^2\gamma^0\, , 
\ee
for charge conjugation. The Dirac field transforms under CPT as
\be
& & \mathcal{CPT} \psi(\hx ) (\mathcal{CPT})^{-1} = CPT \bar{\psi}^{T} ( -x^{\mu},  x^5, x^6 )\, , \quad \mu=0,1,2,3 \, , \nonumber \\
& & (CPT)^{-1} \Gamma^{\mu} CPT = - \Gamma^{\mu\, \dagger}, \quad  (CPT)^{-1} \Gamma^{k } CPT = \Gamma^{k\, \dagger }, \quad k = 5,6 \, ,
\ee
which shows that the extra two-dimensional coordinates have a different CPT transformation property from the ordinary four-dimensional spacetime coordinates. 

Unlike the ordinary four-dimensional spacetime coordinates, the signs of the extra two-dimensional coordinates flip under charge-conjugation $\mathcal{C}$. This is because the pseudoscalar and scalar currents of Dirac spinor are invariant under ordinary charge-conjugation $\mathcal{C}$ in four-dimensional spacetime. If the extra dimensions do not undergo a  flip in sign under charge-conjugation, the action will not be invariant because of the following identities:
\be
& & \mathcal{C} \bar{\psi}(\hx) i\gamma^5  i\partial_5 \psi(\hx) \mathcal{C}^{-1} = i\partial_5\bar{\psi}(\hx)\, i\gamma^5  \psi(\hx)\, , \nonumber \\
& & \mathcal{C} \bar{\psi}(\hx)   i\partial_6 \psi(\hx) \mathcal{C}^{-1} = i\partial_6 \bar{\psi}(\hx)\, \psi(\hx) \, ,
\ee
which have opposite signs compared to the terms imposed by the hermiticity of the action. The same reason applies to the transformation properties of the extra dimensions under  $\mathcal{P}$ and $\mathcal{T}$.

To reflect the intrinsic property of the extra two dimensions, let us introduce an intrinsic W-parity. It is well-known that a massless Dirac spinor can be decomposed into two Weyl spinors in four-dimensional spacetime,
\be \label{WF}
\psi \equiv \psi_L + \psi_R\, , \quad \psi_{L} = \frac{1}{2} (1-\gamma_5) \psi\, , \quad \psi_{R} = \frac{1}{2} (1+\gamma_5) \psi\, .
\ee 
Here $\psi_{L,R}$ are the so-called left-handed and right-handed Weyl spinors with a property
\be
\gamma_5 \psi_{L} = - \psi_L\, , \quad \gamma_5 \psi_{R} = + \psi_R\, .
\ee
They are regarded as two independent Weyl spinors for the massless Dirac spinor in four-dimensional spacetime. However, in the six-dimensional spacetime, the two types of Weyl spinor are correlated via the spin symmetry SP(1,5). It is not difficult to check that the action has an intrinsic discrete symmetry under a W-parity operation $\mathcal{W}$,
\be
& & \mathcal{W} \psi(\hx ) \mathcal{W}^{-1} = W \psi( x^{\mu}, -x^k )\, , \quad k=5,6 \, , \nonumber \\
& &  W^{-1} \Gamma^{\mu} W = -\Gamma^{\mu}, \quad W^{-1} \Gamma^{k} W = \Gamma^{k},  \quad k=5,6, 
\quad W=\Gamma^5 =i\gamma^5 \, .
\ee
Under a combined W-parity and charge-conjugation operation $\mathcal{\tC}\equiv \mathcal{WC}$, we  have
\be
& & \mathcal{\tC} \psi(\hx ) \mathcal{\tC}^{-1} = \tC \bar{\psi}^{T} ( \hx )\, ,  \nonumber \\
& & \tC^{-1} \Gamma^{\hm} \tC = \Gamma^{\hm\, T}\, , \quad \tC= WC= -\gamma^5 \gamma^2\gamma^0 \, ,
\ee
which shows that only under the combined operation $\mathcal{\tC}$ are all six-dimensional coordinates unchanged. Such a combined operator $\mathcal{\tilde{C}}$ has the following feature:  
\be
\left(\psi^{\tc}(\hx)\right)^{\tc} = \mathcal{\tC} \psi^{\tc} (\hx ) \mathcal{\tC}^{-1} = - \psi(\hx) \, ,
\ee
which indicates that the {\it W-parity charge-conjugation} characterizes a discrete $Z_4$ property. 

It is useful to introduce a joint operator $\varTheta \equiv \mathcal{WCPT} = \mathcal{\tC PT}$. Under $\varTheta $ operation, we have
\be
& & \varTheta \psi(\hx ) \varTheta^{-1} = \Theta \bar{\psi}^{T} ( -\hx) \, , \nonumber \\
& & \Theta^{-1} \Gamma^{\hm} \Theta =  \Gamma^{\hm \, \dagger},  \, \quad \Theta = WCPT= \tC PT =\gamma^0\, ,
\ee
which demonstrates that the joint operation $\varTheta $ becomes more essential than the ordinary joint operation CPT for the massless Dirac spinor as a field of six-dimensional spacetime coordinates.

\section{ Massless Dirac spinor as Majorana- or Weyl-Type spinor in 6D Spacetime and Charge Spin Gauge Symmetry}

\subsection{Massless Dirac spinor as Majorana-type spinor in 6D spacetime}

In terms of the Dirac field $\psi^{\tc}(\hx)$ defined via the combined W-parity charge-conjugation $\mathcal{\tilde{C}}$, i.e., 
\be
\psi^{\tc}(\hx) \equiv \mathcal{\tC} \psi(\hx ) \mathcal{\tC}^{-1} = \tC \bar{\psi}^{T} ( \hx ) \, ,
\ee
the action for the massless Dirac spinor in the six-dimensional spacetime can be rewritten as follows
 \be \label{Dirac6DS}
S_{M}^{6d} = \int d^6 x\,  \frac{1}{2} \left( \bar{\psi}(\hat{x}) \Gamma^{\hat{\mu}} i \partial_{\hat{\mu}} \psi(\hat{x}) + 
\bar{\psi}^{\tc} (\hat{x}) \Gamma^{\hat{\mu}} i \partial_{\hat{\mu}} \psi^{\tc}(\hat{x})  \right) 
 \equiv  \int d^6 x\, \frac{1}{2} \bar{\Psi}(\hx) \Gamma^{\hat{\mu}} i \partial_{\hat{\mu}} \Psi(\hx)  \, .
\ee
Here, $\Psi(\hx)$ is an eight-component entity spinor field defined as
\be
\Psi(\hx) \equiv \binom{\psi(\hx) }{\psi^{\tc}(\hx) }\, ,
\ee
which is a Majorana-type spinor in the six-dimensional spacetime, 
\be
\Psi^{\hc}(\hx) =  \mathcal{\hC} \Psi(\hx) \mathcal{\hC}^{-1} = \hC  \bar{\Psi}^{T} ( \hx ) = \Psi(\hx)\, ,
\ee
with $\hC$ given explicitly by
\be
\hC = - i \sigma_{2} \otimes \tC \, .
\ee
Here $\sigma_2$ is the antisymmetry Pauli matrix. The $8 \times 8$ matrix $\hC$ valued in the spinor representation defines a new charge-conjugation in the six-dimensional spacetime. 

It is seen that the W-parity charge-conjugated Dirac field $\psi^{\tc}(\hx)$ enables us to express the complex four-component entity Dirac field $\psi(\hx)$ as an eight-component entity Majorana-type spinor field $\Psi(\hx)$ and obtain a self hermitian action. Considering $\psi(\hx)$ and $\psi^{\tc}(\hx)$ as a {\it charge spin doublet}, we can show that the action given in Eq.~(\ref{Dirac6DS}) possesses an internal {\it charge spin} symmetry SU(2) that characterizes the coherence between $\psi(\hx)$ and $\psi^{\tc}(\hx)$. Namely, the action is invariant under the symmetry group SU(2) transformation 
\be
\Psi(\hx) \to \Psi'(\hx) = e^{\alpha_i \tau_i/2} \Psi(\hx) \, , \quad  (i=1,2,3) \, 
\ee
with $\tau_i/2$ the generators of SU(2).

\subsection{ Charge spin gauge symmetry and doubly electron-charged bosons}

By gauging the {\it charge spin} symmetry SU(2) and introducing the corresponding gauge field $A_{\mu}(\hx) \equiv  g_c A_{\mu}^{i}(\hx)\tau_i /2$, we obtain an action with the internal charge spin gauge symmetry SU(2) for the massless Dirac spinor.  The action is explicitly given as follows
\be \label{Dirac6DG}
S^{6d}_M & = &  \int d^6 x\,  \{ \, \frac{1}{2}\varphi^2(\hx)\, \bar{\Psi}(\hx) \Gamma^{\hat{\mu}} i D_{\hat{\mu}} \Psi(\hx)  
- \frac{1}{2g_c^2} \varphi^2(\hx)\, Tr F_{\hm\hn}(\hx) F^{\hm\hn}(\hx) \nonumber \\
 & + & \frac{1}{2} \varphi^2\partial_{\hm}\varphi(\hx) \partial^{\hm}\varphi(\hx)  - \lambda_s  \frac{1}{6} \varphi^6(\hx) \, \}\, ,
\ee
with $g_c$ and $\lambda_s$  the coupling constants. $D_{\hm}$ and $F_{\hm\hn}(\hx)$ are the covariant derivative and field strength, respectively, in the six-dimensional spacetime: 
\be
& &  iD_{\hm} = i\partial_{\hm} + A_{\hm}(\hx)\, , \nn\\
& &  F_{\hm\hn}(\hx) = i [ D_{\hm}  \,  D_{\hn}] = \partial_{\hm}A_{\hn}(\hx) - \partial_{\hn}A_{\hm}(\hx) - i [A_{\hm}(\hx)\; A_{\hn}(\hx)] \,  .
\ee

When such a Dirac spinor is regarded as a massless charged particle, the third component gauge field $A_{\mu}^3\tau_3/2$ that is associated with the symmetry of subgroup U(1)$\in$SU(2) will characterize an electromagnetic interaction. The gauge bosons in the coset SU(2)/U(1) are doubly electrically charged bosons.

A singlet scalar field $\varphi(\hx)$  is introduced to preserve the global conformal scaling symmetry of the action under the transformations
\be
& & x^{\hm} \to x^{'\hm} = \lambda^{-1}\; x^{\hm}\, ,  \qquad \Psi(\hx) \to \Psi'(\hx') = \lambda^{3/2}\, \Psi(\hx)\, , \nonumber \\ 
& &  A_{\hm} \to A'_{\hm}(\hx') = \lambda\, A_{\hm}(\hx)\, ,\quad \varphi(\hx)\to \varphi'(\hx') = \lambda\, \varphi(\hx)\, .
\ee

From the above action, Eq.~(\ref{Dirac6DG}), we arrive at an equation of motion
\be \label{DiracEQ6D}
& & \Gamma^{\hat{\mu}} i D_{\hat{\mu}} \Psi(\hx) = 0 \, , 
\ee
 for the massless Dirac spinor in the six-dimensional spacetime. In terms of the quadratic form of the covariant derivation, we have 
\be
& & \eta^{\hm\hn}D_{\hm}D_{\hn} \Psi(\hx) = \Sigma^{\hm\hn} F_{\hm\hn}  \Psi(\hx) \, , \qquad \eta^{\hm\hn} = \diag (1, -1, -1, -1, -1, -1)\, .
\ee
For a comparison with the four-dimensional theory, it is useful to rewrite the above equation as follows:
\be
D_{\mu}D^{\mu} \Psi(\hx) - \frac{1}{2}\sigma^{\mu\nu} F_{\mu\nu}  \Psi(\hx) = -\left(\gamma^{\mu}\gamma^5 F_{\mu 5} - i\gamma^{\mu} F_{\mu 6} +\gamma^5 F_{56} \right) \Psi(\hx) - D^{\alpha}D_{\alpha}  \Psi(\hx)  \, ,
\ee
where the right-hand side shows the effect arising from extra dimensions. We have used the following definitions
\be
& & D_{\mu} = \partial_{\mu} - ig_c A_{\mu}^{i} \frac{1}{2} \tau_i\, , \quad D_{\alpha} = \partial_{\alpha} -  ig_c A_{\alpha}^{i} \frac{1}{2} \tau_i\, \quad \alpha = 5,6 \nonumber \\
& & F_{\mu \alpha} = D_{\mu} A_{\alpha}-D_{\alpha} A_{\mu}\, , \quad  \alpha = 5,6\, , \quad F_{56} = D_{5} A_{6}-D_{6} A_{5}\, .
\ee

The equation of motion for the scalar field reads
\be
\partial_{\hm}\partial^{\hm}\varphi^2(\hx)  - 2\lambda_s \varphi^4(\hx) = - F_{\hm\hn}^i F^{\hm\hn\, i} + 2\bar{\Psi}(\hx) \Gamma^{\hat{\mu}} i D_{\hat{\mu}} \Psi(\hx) \, .
\ee 
For the gauge field, we obtain the following equation of motion 
\be
D_{\hn} F^{\hm\hn\, i} (\hx) +  F^{\hm\hn\, i} (\hx) \partial_{\hn}\ln \varphi^2(\hx) = \frac{1}{2} g_c \bar{\Psi}(\hx) \Gamma^{\hm} \frac{1}{2} \tau^i \Psi(\hx)\, .
\ee

So far we have shown that a complex four-component entity massless Dirac spinor in four dimensional spacetime can be realized as an eight-component entity massless Majorana-type spinor in six-dimensional spacetime. The action is explicitly constructed to have the internal charge spin gauge symmetry SU(2) and the maximal global internal spin symmetry SP(1,5) that transforms coherently with the maximal global external Lorentz symmetry SO(1,5).  

In general, the complex Dirac field can be decomposed into real- and imaginary-type spinor fields
\be
& & \psi(\hx) = \psi_{c+}(\hx)  +   \psi_{c-}(\hx)\, , \nonumber \\
& & \psi_{c\pm}(\hx)  \equiv  \frac{1}{\sqrt{2}} ( \psi(\hx) \pm \psi_c(\hx) ) \, , \quad \psi_c(\hx) \equiv C\bar{\psi}^T(\hx) = C\gamma_0 \psi^{\ast}(\hx)
\ee
which satisfy the conditions
\be
C\bar{\psi}_{c+}^T(\hx)  =  \psi_{c+}(\hx) \, , \qquad C\bar{\psi}_{c-}^T(\hx)  =  - \psi_{c-}(\hx) \, .
\ee

The Majorana-type field in the 6D spacetime can be rewritten as 
\be
\Psi(\hx) \equiv \binom{\psi(\hx) }{\psi^{\tc}(\hx) } = \frac{1}{\sqrt{2}} 
 \begin{pmatrix} 1 & 1 \\
                         i\gamma_5 & i\gamma_5 
                         \end{pmatrix}  \Psi_1(\hx)\, , \quad \Psi_1(\hx) \equiv \binom{\psi_{c+}(\hx) }{\psi_{c-}(\hx) } \, .
\ee
The charge conjugation for $\Psi_1(\hx)$ is defined as 
\be
\Psi_1^{\hc}(\hx) =  \mathcal{\hC} \Psi_1(\hx) \mathcal{\hC}^{-1} = \hC  \bar{\Psi}^{T} ( \hx ) = \Psi_1(\hx)\, ,\quad
\hC = \sigma_3\times C
\ee
In terms of $\Psi_1(\hx)$, the action for the spinor field can be expressed as 
\be \label{Dirac6DF1}
S^{6d}_F & = &  \int d^6 x\, \frac{1}{2}\varphi^2(\hx)\, \bar{\Psi}_1(\hx) \Gamma^{\hm} i \tilde{D}_{\hm} \Psi_1(\hx) \, , 
\ee
with 
\be
& & \Gamma^{\hm} = (\gamma^{\mu}, \Gamma^5, \Gamma^6)\, , \quad \Gamma^5 = i \sigma_1\times \gamma^5\, , \; \; \Gamma^6 = \sigma_1 \times I_4 \, ,  \nonumber \\
& & i \tilde{D}_{\hm} = i \partial_{\hm}  + A_{\hm}^i \tau_i/2\, , \quad \tau_1= \sigma_1\times I_4\, , \quad \tau_{i} = \sigma_i \times \gamma_5\; \; (i=2,3).
\ee

In other alternative representations, we have 
\be \label{Dirac6DF2}
S^{6d}_F & = &  \int d^6 x\, \frac{1}{2}\varphi^2(\hx)\, \bar{\Psi}_2(\hx) \Gamma^{\hm} i \tilde{D}_{\hm} \Psi_2(\hx) \, , 
\quad \Psi_2(\hx) \equiv \binom{\psi_{c+}(\hx) }{i\psi_{c-}(\hx) }
\ee
with 
\be
& & \Gamma^{\hm} = (\gamma^{\mu}, \Gamma^5, \Gamma^6)\, , \quad \Gamma^5 = i \sigma_2\times \gamma^5\, , \; \; \Gamma^6 = \sigma_2 \times I_4 \, , \quad  \hC = I_2 \times C  \, , \nonumber \\
& & i \tilde{D}_{\hm} = i \partial_{\hm}  + A_{\hm}^i \tau_i/2\, , \quad \tau_2 = \sigma_2\times I_4\, , \quad \tau_{i} = \sigma_i \times \gamma_5 \; \; (i=1,3).
\ee
and 
\be \label{Dirac6DF3}
S^{6d}_F & = &  \int d^6 x\, \frac{1}{2}\varphi^2(\hx)\, \bar{\Psi}_3(\hx) \Gamma^{\hm} i \tilde{D}_{\hm} \Psi_3(\hx) \, , 
\quad \Psi_3(\hx) \equiv \binom{\psi(\hx) }{\psi_{c}(\hx) }
\ee
with 
\be
& & \Gamma^{\hm} = (\gamma^{\mu}, \Gamma^5, \Gamma^6)\, , \quad \Gamma^5 = i \sigma_3\times \gamma^5\, , \; \; \Gamma^6 = \sigma_3 \times I_4 \, , \quad \hC = \sigma_1\times C \, , \nonumber \\
& & i \tilde{D}_{\hm} = i \partial_{\hm}  + A_{\hm}^i \tau_i/2\, , \quad \tau_3 = \sigma_3\times I_4\, , \quad \tau_{i} = \sigma_i \times \gamma_5 \; \; (i=1,2).
\ee

Different spinor structures reflect relevant properties of extra dimensions and representations of internal symmetry.

\subsection{Majorana-Weyl property of massless Dirac spinor }

To reflect explicitly the properties of chirality spin and W-parity, it is useful to show that a complex four-component entity massless Dirac spinor in four dimensional spacetime can be realized as an eight-component entity massless Weyl spinor in six-dimensional spacetime as follows
 \be \label{Dirac6DW}
S_{W}^{6d} = \int d^6 x\,  \frac{1}{2} \left( \bar{\psi}_{-}(\hx) \Gamma^{\hat{\mu}} i \partial_{\hat{\mu}} \psi_{-}(\hx) + 
H.c.  \right)  \, ,
\ee
with 
\be
\Gamma^{\hm} = (\gamma^{\mu}, \Gamma^5, \Gamma^6)\, , \quad \Gamma^5 = i\sigma_1 \otimes \gamma^5\, , \quad 
\Gamma^6 = i\sigma_2 \otimes \gamma^5\, .
\ee
$\psi_{-}(\hx)$ is an eight-component entity massless Weyl-type spinor defined as
\be \label{WF6D}
& & \psi_{-}(\hx)\equiv \binom{\psi_L(\hx)}{\psi_R(\hx)} \, , \quad 
\gamma_7\, \psi_{-}(\hx) = - \psi_{-}(\hx) \, , \quad \gamma_7 = \sigma_3 \otimes \gamma_5 
\ee
with $\psi_L(\hx)$ and $\psi_R(\hx)$ the left-handed and right-handed Weyl spinors defined in Eq.~(\ref{WF}).  It is clear that the action Eq.~(\ref{Dirac6DW}) is invariant under the Lorentz symmetry group SO(1,5) and spin symmetry group SP(1,5)$\cong$SO(1,5) with generators and constant metric matrix
\be
& & \Sigma_{\hm\hn} = \frac{i}{4} [ \Gamma_{\hm},  \Gamma_{\hn}]\, , \nn \\
& & \eta_{\hm\hn} = \frac{1}{2}\{  \Gamma_{\hm},  \Gamma_{\hn} \}  \, , \qquad  \eta_{\hm\hn} = \diag.(1,-1,-1,-1,-1,-1)\, .
\ee
It is easy to show that both fifth and sixth dimensions are spatial under operations ${\mathcal P}$ and ${\mathcal T}$. We can define the charge conjugation for such a Weyl-type spinor in the six-dimensional spacetime. Explicitly, the charge-conjugated Weyl spinor $\psi_{-}^{c}(\hx) $ is defined as
 \be
 & & \psi_{-}^{\bar{c}}(\hx) \equiv \mathcal{C}_6 \psi_{-}(\hx) \mathcal{C}_6^{-1} = C_6 \bar{\psi}_{-}^T(\hx) =  \binom{-i\psi_R^{\; c}(\hx)}{i\psi_L^{\; c} (\hx)} \, , 
\ee
with the property for the charge-conjugation operator 
\be
& & C_6 \Gamma^{\hm} C_6^{-1} = - \Gamma^{\hm\, T}\, , \quad C_6 = \sigma_2\otimes C\, , \quad C= i\gamma_2\gamma_0 \, , \nonumber \\
& &  \left(\psi_{-}^{\bar{c}}(\hx) \right)^{\bar{c}} \equiv \mathcal{C}_6 \psi_{-}^{\bar{c}} (\hx) \mathcal{C}_6^{-1}  = -  \psi_{-}(\hx)\, .
\ee

The action Eq.~(\ref{Dirac6DW}) can be rewritten as follows:
\be \label{Dirac6DW2}
S_{W}^{6d} & = &\int d^6 x\,  \frac{1}{2} \left( \bar{\psi}_{-}(\hx) \Gamma^{\hat{\mu}} i \partial_{\hat{\mu}} \psi_{-}(\hx) + 
\bar{\psi}_{-}^{\bar{c}} (\hx) \Gamma^{\hat{\mu}} i \partial_{\hat{\mu}} \psi_{-}^{\bar{c}}(\hx)  \right)  \nn \\
& \equiv &  \int d^6 x\, \frac{1}{2} \bar{\Psi}_{-}(\hx) \Gamma^{\hat{\mu}} i \partial_{\hat{\mu}} \Psi_{-}  \, ,
\ee
where we have taken $\psi_{-}(\hx)$ and $\psi_{-}^{c}(\hx) $ as a {\it charge spin doublet} to define the following sixteen-component entity spinor field
\be \label{MWF1}
\Psi_{-}(\hx) \equiv \binom{\psi_{-}(\hx) }{\psi_{-}^{\bar{c}}(\hx) }\, . 
\ee
We can check that the action Eq.~(\ref{Dirac6DW2}) is equivalent to the action Eq.~(\ref{Dirac6DS}) by noticing the following identity
\be
\psi_{-}^{{\bar{c}}}(\hx) = \binom{\psi_R^{\; \tc}(\hx)}{\psi_L^{\; \tc}(\hx)} \, .
\ee 
It is not difficult to show that $\Psi_{-}(\hx)$  satisfies a Majorana-type condition 
\be \label{MWF2}
& & \Psi_{-}^{c} (\hx) = \mathcal{C}_8 \Psi_{-} (\hx) \mathcal{C}_8^{-1}  = \Psi_{-}(\hx) \, , \nonumber \\
& & C_8 = \sigma_2\otimes C_6 = \sigma_2\otimes \sigma_2\otimes C \, .
\ee

The action Eq.~(\ref{Dirac6DW2}) possesses the charge spin symmetry SU(2) between Weyl-type spinor $\psi_{-}(\hx)$ and its charge-conjugation $\psi_{-}^{\bar{c}}(\hx)$. Again taking SU(2) as a gauge symmetry, we can obtain, analogous to the action Eq.~(\ref{Dirac6DG}), the following gauge invariant action
\be \label{Dirac6DMWG}
S^{6d}_{MW} & = &  \int d^6 x\,  \{ \frac{1}{2} \varphi^2(\hx)\bar{\Psi}_{-}(\hx) \Gamma^{\hat{\mu}} i D_{\hat{\mu}} \Psi_{-}(\hx) \nonumber \\
& - & \frac{1}{4} \varphi^2(\hx) F_{\hm\hn}^i(\hx) F^{\hm\hn\, i}(\hx)  + \frac{1}{2}\varphi^2(\hx) \partial_{\hm}\varphi(\hx) \partial^{\hm}\varphi(\hx) - \frac{1}{6} \lambda_s \varphi^6(\hx)  \} \, .
\ee
Here $\Psi_{-}(\hx)$ is regarded as a Majorana-Weyl-type spinor in the spinor representation of eight dimensions. 

The extra dimensions correlate with the chirality spin of the Dirac spinor. In general, a massless Dirac spinor is characterized by the intrinsic quantum numbers of boost spin, helicity spin, chirality spin and charge spin.

\section{Gravifield Fiber Bundle Structure of Spacetime and Gravitational Gauge Field Theory in 6D Spacetime }
 
The above action for the massless Dirac spinor with maximal symmetry is built based on a {globally flat Minkowski spacetime }, which is an {\it affine spacetime} denoted as $M_A$ for a convenience. It possesses in general a Poincar\'e or non-homogeneous Lorentz symmetry P(1,5) = SO(1,5)$\ltimes P^{1,5}$. Both internal spin symmetry SP(1,5) of the massless Dirac spinor and external Lorentz symmetry SO(1,5) of the coordinates are global symmetries. They have to coherently incorporate each other to preserve the Lorentz invariance of the action in the six-dimensional spacetime. In this section,  we shall propose that the spinor spin symmetry SP(1,5), analogous to other internal symmetries of spinors, is gauged as a local symmetry,  and the external Lorentz symmetry SO(1,5) remains a global symmetry. Thus the spinor spin symmetry SP(1,5) as an internal symmetry is distinguished from the external Lorentz symmetry SO(1,5).  To build an action with both the local spin gauge symmetry SP(1,5) and the global Lorentz symmetry SO(1,5),  we shall apply the postulate proposed in Ref.~\cite{YLWU} to construct an action within the framework of {\it gravitational quantum field theory}.

\subsection{Gravifield fiber bundle structure of spacetime}

As demonstrated in Ref.~\cite{YLWU}, it is essential to introduce a {\it bicovariant vector field} and a spin gauge field to preserve both the local spin gauge symmetry SP(1,5) and the global Lorentz symmetry SO(1,5). Explicitly, the kinematic term for the Dirac spinor field is extended to be  
\be
\Gamma^{\hm}i\partial_{\hm} \to \Gamma^{\ha}\chih_{\ha}^{\;\, \hm}(\hx) \left(i\partial_{\hm} + g_s {\mathcal A}_{\hm}^{\; \hb\ch}(\hx)\frac{1}{2}\Sigma_{\hb\ch}  \right) \, ,
\ee
with $\chih_{\ha}^{\;\, \hm}(\hx)$ the {\it bicovariant vector field}, and 
\be
 {\mathcal A}_{\hm}(\hx) = g_s  {\mathcal A}_{\hm}^{\; \hb\ch}(\hx)\, \frac{1}{2}\Sigma_{\hb\ch} \, ,
\ee
the spin gauge field. Here the Greek alphabet ($\hm, \hm = 0,1,2,3,5,6$) and the Latin alphabet ($\ha, \hb,=0,1,2,3,5,6$)  are adopted to distinguish the vector indices defined in the vector representations of Lorentz group SO(1,5) and spin group SP(1,5), respectively. 

The derivative vector operator $\partial_{\hm} \equiv \partial/\partial x^{\hm}$ at the point $\hx$ of $M_A$ defines a {\it tangent basis } $\{\partial_{\hm}\}\equiv \{\partial/\partial x^{\hm}\} $ for the {\it tangent Minkowski spacetime} $T_M$ over the globally flat Minkowski spacetime $M_A$. Accordingly, we introduce a {\it field vector} $\chih_{\ha}(\hx)$ at point $\hx$ of $M_A$ respective to the derivative vector operator $\partial_{\hm}$.  Such a field vector $\chih_{\ha}(\hx)$ is explicitly defined via the bi-covariant vector field $\chih_{\ha}^{\;\, \hm}(\hx)$ as follows: 
\be
\chih_{\ha}(\hx) \equiv \chih_{\ha}^{\;\, \hm}(\hx) \partial_{\hm}\, ,
\ee
which forms a field basis $ \{\chih_{\ha}(\hx) \} $ for the locally flat non-coordinate spacetime over the globally flat Minkowski spacetime $M_A$. We shall call such a locally flat noncoordinate spacetime a {\it gravifield spacetime} denoted as $G_M$. Here $\chih_{\ha}^{\;\, \hm}(\hx)$ is the so-called {\it gravifield} and $ \{\chih_{\ha}(\hx) \} $ provides a {\it gravifield basis}.

The displacement vector $dx^{\mu}$ at point $\hx$ of $M_A$ defines a {\it dual tangent basis} $\{dx^{\hm}\}$ for a {\it dual tangent Minkowski spacetime} $T_M^{\ast}$ over the globally flat Minkowski spacetime $M_A$. The tangent basis and dual tangent basis satisfy the dual condition
\be
 < dx^{\hm},\, \partial/\partial x^{\hn}  > = \frac{\partial x^{\hm}}{\partial x^{\hn}} = \eta_{\hn}^{\; \hm}\, .
\ee

Analogously, we shall introduce a {\it dual vector} $\chi^{\, \ha}(\hx)$ at  point $\hx$ of $M_A$ respective to the displacement vector $dx^{\mu}$.  For that, let us first define a {\it dual bicovariant vector field} $\chi_{\hm}^{\;\; \ha}(\hx)$ via the following orthonormal conditions  
\begin{eqnarray}
\chi_{\hm}^{\;\; \ha} (\hx)\, \chih^{\;\; \hm}_{\hb}(\hx)   =  \chi_{\hm}^{\;\; \ha} (\hx) \chih_{\hb\, \hn}(\hx) \eta^{\hm\hn} = \eta^{\;\; \ha}_{\hb} \; , \quad \chi_{\hm}^{\;\; \ha}(\hx) \chih^{\;\;\hn}_{\ha}(\hx) = \chi_{\hm\, \ha} (\hx) \chih_{\hb}^{\;\; \hn}(\hx)  \eta^{\ha\hb} = \eta_{\hm}^{\;\;\hn} \, ,
\end{eqnarray}
which can be regarded as the inverse of the gravifield $\chih_{\ha}^{\;\;\hm}(\hx)$. Thus the dual bicovariant vector $\chi_{\hm}^{\;\, \ha}(\hx)$ is dual to  the gravifield, which exists once the determinant of $\chih_{\ha}^{\;\; \hm}(\hx)$ is nonzero, $\det \chih_{\ha}^{\;\; \hm}(\hx)\neq 0$. 

The {\it dual vector} $\chi^{\, \ha}(\hx)$ is defined via the {\it gauge-type gravifield} $\chi_{\hm}^{\;\, \ha}(\hx)$ associated with the displacement vector $dx^{\mu}$ 
\be
\chi^{\; \ha} (\hx)  = \chi_{\hm}^{\;\; \ha} (\hx) dx^{\hm}  
\ee
which satisfies the dual condition 
 \be
 < \chi^{\, \ha},\,  \chih_{\hb} > = \chi_{\hm}^{\;\; \ha}(\hx)  \chih_{\hb}^{\;\; \hn} (\hx)  < dx^{\hm} ,\, \partial_{\hn}> =  \chi_{\hm}^{\;\; \ha}(\hx)  \chih_{\hb}^{\;\; \hn} (\hx)  \eta_{\hn}^{\; \hm} = \eta_{\hb}^{\;\, \ha} \, .
\ee
The {\it dual gravifield basis} $\{\chi^{\, \ha} \}$ forms a {\it dual gravifield spacetime} $G_M^{\ast}$ over the globally flat Minkowski space-time $M_A$.

The gravifield $\hat{\chi}_{\ha}^{\;\;\hm}(\hx)$ defined on the gravifield spacetime $G_M$ and valued on the tangent Minkowski spacetime $T_M$ transforms as a bicovariant vector field under both the local spin gauge transformation SP(1,5) and global Lorentz transformation SO(1,5). Such a gravifield basis does not commute and satisfies the following non-commutation relation 
\be
& & [ \chih_{\ha} ,\; \chih_{\hb}] = f_{\ha\hb}^{\; \hc}\,  \chih_{\ch}, \qquad  f_{\ha\hb}^{\; \ch} \equiv - \frac{1}{2} \left( \chih_{\ha}^{\;\; \hm} \chih_{\hb}^{\;\; \hn} - \chih_{\hb}^{\;\; \hm} \chih_{\ha}^{\;\; \hn} \right) \chi_{\hm\hn}^{\; \ch} \, , \nonumber \\
& &  \chi_{\hm\hn}^{\; \ch}  = \partial_{\hm}\chi_{\hn}^{\;\; \ch} - \partial_{\hn}\chi_{\hm}^{\;\; \ch} \, ,
\ee 
which shows that the locally flat gravifield spacetime $G_M$ is associated with a non-commutative geometry. Such a non-commutative geometry is characterized by a gravitational field strength $\chi_{\hm\hn}^{\; \ha} $ defined from the {\it gauge-type gravifield} $\chi_{\hm}^{\;\; \ha}(\hx)$. We write such a {\it gauge-type gravifield} as follows
\be
\dg_{\hm}(\hx) = \chi_{\hm}^{\;\; \ha}(\hx) \frac{1}{2} \Gamma_{\ha} ,
\ee
which is defined in the dual tangent Minkowski spacetime $T_M^{\ast}$ and valued on the dual gravifield spacetime $G_M^{\ast}$. 

Geometrically, we arrive at a biframe spacetime $T_M\times G_M$ associated with its dual spacetime $T_M^{\ast}\times G_M^{\ast}$ over the spacetime $M_A$. Based on the nature of globally and locally flat vector spacetimes,  it allows for the canonical identification of vectors in tangent Minkowski spacetime $T_M$ at points with vectors (points) in Minkowski spacetime itself $M_A$, and also for the canonical identification of vectors at a point with its dual vectors at the same point. 

Physically, the globally flat Minkowski spacetime is deemed a {\it vacuum spacetime} $\bf{V}$.  Thus the canonical identification of vector spacetimes enables us to simplify the spacetime structure as
\be
 T_M \cong T_M^{\ast} \cong  M_A & \equiv & \bf{V}\, , \nonumber \\
G_M \cong G_M^{\ast} & \equiv & \bf{G}\, .
\ee
Therefore, the total spacetime is viewed as a {\it gravifield fiber bundle} $\bf{E}$ with the {\it gravifield spacetime} $\bf{G}$ as the fiber and the vacuum spacetime $\bf{V}$ as the base spacetime. The correlation between the total spacetime $\bf{E}$ and the product spacetime $\bf{V\times G}$ is defined using a continuous surjective map $\pi_{\chi}$ which projects the bundle $\bf{E}$ to the base spacetime $\bf{V}$, i.e., $\pi_{\chi}$: $\bf{E \to V}$. Formally, we express the {\it gravifield fiber bundle structure} of spacetime as  $(\bf{E, V, \pi_{\chi}, G}) $ with the trivial case
\be
\bf{E} \sim \bf{V} \times \bf{G}\, .
\ee

\subsection{Gravitational gauge field theory for massless Dirac spinor in 6D spacetime}

With the above analysis, we are in a position to build a gravitational gauge field theory for a massless Dirac spinor based on the framework of gravitational quantum field theory proposed in Ref.~\cite{YLWU}. Our main postulates are that: (i) the maximal internal symmetry of the gravifield spacetime $\bf{G}$ and the Dirac spinor field is a gauge symmetry that characterizes the basic interaction and dynamics of all fields, and the maximal external symmetry of vacuum spacetime $\bf{V}$ is a global symmetry that describes the inertial motion and kinematics of all fields; and (ii) the action is built based on the principles of gauge-invariance and coordinate-independence.  

In terms of the gravifield basis $\{\chi^{\ha}\}$ and $\{\hat{\chi}_{\ha}\}$,  it enables us to define a non-coordinate exterior differential operator in the gravifield spacetime
\be
d_{\chi} =  \chi^{\ha}\wedge \hat{\chi}_{\ha}   \, .
\ee
Thus all gauge fields and field strengths can be expressed as the one-form and two-form gravifield spacetime $\bf{G}$ by using the gravifield basis vector $\chi^{\ha} $ and exterior differential operator. Explicitly, we have  
\be 
& &  {\mathcal A} = -i  {\mathcal A}_{\ha}\, \chi^{\ha}\; , \qquad {\cal F} = d_{\chi} \,  {\mathcal A} +  {\mathcal A}\wedge  {\mathcal A} =\frac{1}{2i} {\cal F}_{\ha\hb}\, \chi^{\ha}\wedge \chi^{\hb} \; , \nonumber \\
& & \dg = -i \dg_{\ha} \, \chi^{\ha}\; , \qquad {\cal G} = d_{\chi} \, \dg +  {\mathcal A}\wedge \dg  + \Wm \wedge \dg = \frac{1}{2i} {\cal G}_{\ha\hb}\, \chi^{\ha}\wedge \chi^{\hb} \; , \nonumber \\
& & A = -i A_{\ha}\, \chi^{\ha}\; , \qquad F = d_{\chi}\, A + A \wedge A = \frac{1}{2i} F_{\ha\hb}\, \chi^{\ha}\wedge \chi^{\hb} \; , \nonumber \\
& & \Wm = -i \Wm_{\ha}\, \chi^{\ha}\; , \qquad {\cal W} = d_{\chi}\, \Wm = \frac{1}{2i} {\cal W}_{\ha\hb}\, \chi^{\ha}\wedge \chi^{\hb} \; ,
\ee
where the gauge fields and field strengths are all sided on the locally flat gravifield spacetime $\bf{G}$. They are projected through the gravifield $\hat{\chi}_{\ha}^{\;\; \hm}$ to become the corresponding gauge fields and field strengths defined in the globally flat vacuum spacetime $\bf{V}$, i.e.,
\be
 & & {\bf{A}}_{\ha} = \hat{\chi}_{\ha}^{\;\; \hm}(\hx) {\bf{A}}_{\hm}(\hx) \; , \qquad {\bf{F}}_{\ha\hb} =  \hat{\chi}_{\ha}^{\;\; \hm}(\hx)\,\hat{\chi}_{\hb}^{\;\; \hn}(\hx) {\bf{F}}_{\hm\hn}(\hx) \, , 
\ee
with ${\bf{A}} = (  {\mathcal A},\, \dg,\, A,\, \Wm )$ and ${\bf F} = ( {\cal F},\, {\cal G},\, F, \, {\cal W} )$.
The field strength of the gauge-type gravifield $\mathbf{G}_{\hm}(\hx)$ has the following explicit form 
\be
{\cal G}_{\hm\hn}(\hx) & = & \nabla_{\hm}\dg_{\hn}  - \nabla_{\hn}\dg_{\hm} =  [\, \nabla_{\hm}\chi_{\hn}^{\;\; \ha} (\hx)  - \nabla_{\hn}\chi_{\hm}^{\;\; \ha}(\hx) \,] \, \frac{1}{2}\Gamma_{\ha} \equiv {\cal G}_{\hm\hn}^{\ha}(\hx) \, \frac{1}{2}\Gamma_{\ha}  \; , \nonumber \\
{\cal G}_{\hm\hn}^{\ha}(\hx)  & = & (\partial_{\hm} + g_w W_{\hm} ) \chi_{\hn}^{\;\; \ha}  + g_s {\mathcal A}_{\hm\;\; \hb}^{\; \ha}  \chi_{\hn}^{\;\; \hb}  - (\partial_{\hn}+ g_w W_{\hn} )\chi_{\hm}^{\;\; \ha}   - g_s {\mathcal A}_{\hn\;\; \hb}^{\; \ha}  \chi_{\hm}^{\;\; \hb} \, ,
\ee
with the covariant derivative defined as 
\be
i\nabla_{\hm} =i \partial_{\hm}  +  {\mathcal A}_{\hm} + \Wm_{\hm}  = i\bigtriangledown_{\hm}  + \Wm_{\hm} \, , 
\quad i\btd_{\hm} = i\partial_{\hm}  +  {\mathcal A}_{\hm} \, .
\ee
Here the gauge field $\Wm_{\hm}= g_w W_{\hm}$ with gauge coupling constant $g_w$ is introduced to extend the {\it global conformal scaling symmetry} of the massless Dirac spinor to a local {\it conformal scaling gauge symmetry}, while allowing the coordinates of Minkowski spacetime to remain, keeping the global conformal scaling symmetry. Namely, the Dirac spinor field, scalar field and gravifield transform under the local {\it conformal scaling gauge transformation} as follows: 
\begin{eqnarray}
& & \Psi(\hx) \to \Psi'(\hx) = \xi^{3/2} (\hx) \Psi(\hx) \; , \qquad  \varphi(\hx) \to \varphi'(\hx) = \xi (\hx) \varphi(\hx)\, , \nonumber \\
& & \hat{\chi}_{\ha}^{\;\; \hm}(\hx) \to  \hat{\chi}_{\ha}^{'\; \hm}(\hx) = \xi(\hx) \hat{\chi}_{\ha}^{\;\; \hm}(\hx) \, , \qquad \chi_{\hm}^{\;\; \ha}(\hx) \to \chi_{\hm}^{'\; \ha}(\hx) = \xi^{-1}(\hx)  \chi_{\hm}^{\;\; \ha}(\hx) \, ,
\end{eqnarray}
and the {\it conformal scaling gauge field} transforms as an Abelian gauge field,
\begin{eqnarray}
 \Wm_{\hm}(\hx) \to \Wm'_{\hm}(\hx) = \Wm_{\hm}(\hx) + \partial_{\hm} \ln \xi(\hx)\, .
\end{eqnarray}
Thus the invariant field strength is given by
\be
{\cal W}_{\hm\hn} = \partial_{\hm}\Wm_{\hn} - \partial_{\hn}\Wm_{\hm} \, ,
\ee
which governs a basic force of {\it conformal scaling gauge interaction}. Such a conformal scaling gauge field was first proposed by Weyl~\cite{WG} for the purpose of the electromagnetic field, which is known to result from an U(1) gauge symmetry.  

It is useful to express the covariant derivative as one-form in the gravifield spacetime ${\bf G}$ 
\be
& & {\mathcal D} \equiv \chi^{\ha} {\mathcal D}_{\ha}  =  \chi^{\ha} \left( \hat{\chi}_{\ha} -i A_{\ha} - i  {\mathcal A}_{\ha} \right)\, , \qquad {\mathcal D}_{\ha} = \hat{\chi}_{\ha}^{\;\;\hm} {\mathcal D}_{\hm} = \hat{\chi}_{\ha}^{\;\;\hm} (\, \partial_{\hm} -i A_{\hm} -i  {\mathcal A}_{\mu}  \, )  \, .
\ee
The Hodge star $``\ast"$ in the six-dimensional gravifield spacetime ${\bf G}$ is defined as 
\be
\ast {\bf F} = \frac{1}{2!4! 2i} \epsilon_{\ha\hb\ch\hd\he\hf}\, \eta^{\ha\ha'}\eta^{\hb\hb'} {\bf F}_{\ha'\hb'}\, \chi^{\ch}\wedge \chi^{\hd} \wedge \chi^{\he}\wedge \chi^{\hf} .
\ee

With the exterior differential operator $ d_{\chi}$ and the gravifield basis vectors $\chi^{\ha}$ and $\hat{\chi}_{\ha}$, it enables us to build a gauge-invariant and coordinate-independent action  in the gravifield spacetime ${\bf G}$. The general form of the action is found to be 
\be
\label{Dirac6DGG}
S_{\chi}^{6d}  & = & \int \{\, \frac{1}{2}\varphi^2\, [\, i \bar{\Psi}_{-}\, \dg \wedge \ast {\mathcal D} \, \Psi_{-}\,  
-  \frac{1}{g_c^2}\, Tr\, F \wedge \ast F  - \frac{1}{2g_w^2}\, {\cal W} \wedge \ast {\cal W}  - \frac{2}{3g_s^2} \, Tr\, {\cal F} \wedge \ast {\cal F}  \, ]    \nonumber \\
& + & \frac{1}{12}\varphi^4 \, [\, \alpha_W \, Tr\, {\cal G}\wedge \ast {\cal G}  - 2\alpha_E \,  Tr\, {\cal F} \wedge \ast (\dg \wedge \dg ) \, ]  \nonumber  \\
& - & \frac{1}{2} \varphi^2 d \varphi \wedge \ast d \varphi  + \frac{1}{12} \lambda_s \varphi^6  \, Tr\, (\dg \wedge \dg)\wedge \ast (\dg \wedge \dg )\,   \} \, ,
\ee
with $\alpha_E$ and $\alpha_W$ the constant parameters.  The following definitions and relations have been used
\be
& & \dg \equiv \eta_{\ha}^{\; \hb} \frac{1}{2}\Gamma_{\hb}\, \chi^{\ha} \, , \qquad \ast \chi^{\ha} = \frac{1}{5!} \, \epsilon^{\ha}_{\;\; \hb\ch\hd\he\hf} \,  \chi^{\hb} \wedge \chi^{\ch} \wedge \chi^{\hd} \wedge \chi^{\he} \wedge \chi^{\hf} \; , \nonumber \\
& & (\dg \wedge \dg )  \equiv \eta_{\ha}^{\;\ha'} \eta_{\hb}^{\;\hb'} \, \frac{1}{2i} \Sigma_{\ha'\hb'}\, \chi^{\ha} \wedge \chi^{\hb} \; , \quad \ast (\dg \wedge \dg)  = \frac{1}{2!4!}\eta_{\ha}^{\;\ha'} \eta_{\hb}^{\;\hb'} \, \frac{1}{2i} \Sigma_{\ha'\hb'}\, \epsilon^{\ha\hb}_{\;\;\;\; \ch\hd\he\hf}\, \chi^{\ch} \wedge \chi^{\hd} \wedge \chi^{\he} \wedge \chi^{\hf} \; , \nonumber \\
& & d \varphi \equiv  (d_{\chi} - i \Wm )\varphi\; , \qquad  \ast d \varphi =  \frac{1}{5!}\,  (\hat{\chi}_{\ha} - \Wm_{\ha})\varphi  \, \epsilon^{\ha}_{\;\; \hb\ch\hd\he\hf} \,  \chi^{\hb} \wedge \chi^{\ch} \wedge \chi^{\hd} \wedge \chi^{\he} \wedge \chi^{\hf} \; ,
\ee
with $ \epsilon^{\ha\hb\ch\hd\he\hf}$ ( $\epsilon^{012356} = 1$,  $ \epsilon^{\ha\hb\ch\hd\he\hf} = - \epsilon_{\ha\hb\ch\hd\he\hf} $) the totally antisymmetric Levi-Civita tensor which has the following general properties
\be
& &  \epsilon_{a_1...a_n} \epsilon^{b_1...b_n} = - n! \,\eta^{b_1}_{[a_1}...\eta^{b_n}_{a_n]}\, , \qquad 
 \epsilon_{a_1...a_n} \epsilon^{a_1...a_n} = - n!\, ,   \nonumber \\
 & &\epsilon_{a_1...a_k a_{k+1}...a_n} \epsilon^{a_1...a_k b_{k+1}...b_n} = - k! (n-k)!\, \eta^{b_{k+1}}_{[a_{k+1}}...\eta^{b_n}_{a_n]}\, , \nonumber \\
 & & \det \mathsf{M} = \frac{1}{n!}\, \epsilon^{a_1\cdots a_n} \epsilon^{b_1\cdots b_n} M_{a_1 b_1} \cdots M_{a_n b_n}\, ,
 \ee
with $\mathsf{M}$ an $n\times n$ matrix $\mathsf{M} = (M_{ab})$.

The action given in Eq.~(\ref{Dirac6DGG}) provides a gravitational gauge field theory for the massless Dirac spinor with the maximal gauge symmetry group
\be
G_S = SU^{\ast}(4)\times SU(2)\times SG(1)\, .
\ee
Here SG(1) denotes the conformal scaling gauge symmetry. The Dirac spinor field and gauge field belong to the spinor representation and vector representation of the spin gauge group SP(1,5)$\cong$ SU$^{\ast}$(4), respectively, in the six-dimensional gravifield spacetime ${\bf G}$.

\section{ Dynamics of Fields and Spacetime in Gravitational Gauge Field Theory}

The action Eq.~(\ref{Dirac6DGG}) is obtained following the principle of gauge-invariance and coordinate-independence in the locally flat gravifield spacetime ${\bf G}$, which is distinguished from the general theory of relativity which was built based on the principle of general covariance of coordinate with a local symmetry group GL(D, R) in a curved Riemannian spacetime. 

To see explicitly the difference between the gravitational gauge field theory and the general theory of relativity, it is useful to take a formalism by projecting the action Eq.~(\ref{Dirac6DGG}) from the locally flat gravifield spacetime ${\bf G}$ to the gravifield fiber bundle with the globally flat vacuum spacetime ${\bf V}$ as a base spacetime. It can be realized by simply changing the gravifield basis $\{\chi^{\ha}\}$ and $\{\hat{\chi}_{\ha}\}$ into the corresponding coordinate basis $\{ dx^{\hm}\}$ and $\{ \partial_{\hm}\}$. The explicit formalism is found to be
\be
\label{Dirac6DGM}
S_{\chi}^{6d}  & = & \int d^{6}x\; \chi\,  \{\hat{\chi}^{\hm\hn} \bar{\Psi}_{-} \dg_{\hm} i {\mathcal D}_{\hn} \Psi_{-} 
- \frac{1}{4}\phi \,  \hat{\chi}^{\hm\hm'} \hat{\chi}^{\hn\hn'} [\, F^i_{\hm\hn} F^{i}_{\hm'\hn'} 
+ {\cal W}_{\hm\hn} {\cal W}_{\hm'\hn'} + {\cal F}_{\hm\hn}^{\ha\hb} {\cal F}_{\hm'\hn'\ha\hb} \, ] \nonumber \\
& + & \frac{1}{4} \phi^2 \, \hat{\chi}^{\hm\hm'} \hat{\chi}^{\hn\hn'} \, {\cal G}_{\hm\hn}^{\ha} {\cal G}_{\hm'\hn' \ha} - \phi^2 \alpha_E g_s  \hat{\chi}^{\;\;\hm}_{\ha} \hat{\chi}^{\;\;\hn}_{\hb}  {\cal F}_{\hm\hn}^ {\ha \hb}  +  \frac{1}{2}\hat{\chi}^{\hm\hn} d_{\hm} \phi d_{\hn}\phi  - \lambda_s \phi^3  \},  
\ee
with the definitions 
\be \label{tensor}
& & \bar{\Psi}_{-} = \Psi^{T}_{-} C_8 \, , \quad \hat{\chi}^{\hm\hn}(\hx) = \hat{\chi}_{\ha}^{\;\;\hm}(\hx) \hat{\chi}_{\hb}^{\;\;\hn}(\hx) \eta^{\ha\hb}\, , \nonumber \\
& &  \phi \equiv \varphi^2/2\, , \quad d_{\hm} \phi = (\partial_{\mu} - 2 g_w W_{\mu} ) \phi\, ,
\ee
where the symmetric tensor field $ \hat{\chi}^{\mu\nu}(x)$ couples to all fields. We have also made a redefinition by rescaling the Majorana-Weyl type spinor field $\Psi_{-}(\hx)$ Eqs.(\ref{MWF1})-(\ref{MWF2}) to be
\be
 \Psi_{-}(\hx)\to \Psi_{-}(\hx)/\varphi(\hx) = \Psi_{-}(\hx)/ \sqrt{2\phi(\hx)} \, .
\ee

\subsection{Generalized equations of motion in a gravitational relativistic quantum theory}

From the above action, we are able to extend the Dirac equation in  four-dimensional relativistic quantum theory to a generalized equation in the six-dimensional gravitational relativistic quantum theory. Explicitly, a generalized equation of motion for the massless Dirac spinor with maximal symmetry is simply given by
\be \label{EM-F}
& & \Gamma^{\ha} \hat{\chi}_{\ha}^{\;\; \hm} i ( {\mathcal D} _{\hm} + {\mathsf V} _{\hm} ) \Psi_{-}  =   0 \, , 
\ee
with the spin gauge-invariant vector field defined as 
\be
{\mathsf V}_{\hm} (\hx) & \equiv &  \frac{1}{2} \hat{\chi}\chi_{\hm}^{\;\; \hb} \btd_{\hr} (\chi \hat{\chi}_{\hb}^{\;\; \hr})    = \frac{1}{2} (\partial_{\hm}\ln \chi  + g_w W_{\hm}) -\frac{1}{2}\chih_{\hb}^{\;\; \hr}\nabla_{\hr}\chi_{\hm}^{\;\; \hb} \, ,
\ee
which preserves the conformal scaling gauge invariance of the equation of motion.

Its quadratic form is found to be
\be
& & \chih^{\hm\hn} (\nabla_{\hm} + {\mathsf V} _{\hm} ) ( {\mathcal D} _{\hn} + {\mathsf V} _{\hn} )  \Psi_{-} \nonumber \\
& & =   \Sigma^{\ha\hb} \chih_{\ha}^{\;\; \hm} \chih_{\hb}^{\;\; \hn} [\, 
 {\cal F}_{\hm\hn} + F_{\hm\hn} + i {\cal V}_{\hm\hn} - {\cal G}_{\hm\hn}^{\ch} \chih_{\ch}^{\;\; \hr} i ( {\mathcal D} _{\hr} + {\mathsf V} _{\hr} )\, ]  \Psi_{-} \, , 
 \ee
 where we have introduced the definitions
 \be
 & & (\nabla_{\hm} + {\mathsf V} _{\hm} ) ( {\mathcal D} _{\hn}  + {\mathsf V} _{\hn} ) \equiv ({\mathcal D} _{\hm} + {\mathsf V} _{\hm} ) ( {\mathcal D} _{\hn} + {\mathsf V} _{\hn} )  + \Gamma_{(\hm\hn)}^{\hr} ( {\mathcal D} _{\hr} + {\mathsf V} _{\hr} ) \, ,\nonumber \\
 & &  {\cal G}_{\hm\hn}^{\ha}  = \nabla_{\hm}\chi_{\hn}^{\;\; \ha} - \nabla_{\hn}\chi_{\hm}^{\;\; \ha}\, ,  \quad \nabla_{\hm}\chi_{\hn}^{\;\; \ha} = (\partial_{\hm} + g_w W_{\hm} ) \chi_{\hn}^{\;\; \ha} + g_s  {\mathcal A}_{\hm\, \hb}^{\ha}  \chi_{\hn}^{\;\;\hb}\, , \nonumber \\
 & &  {\cal V}_{\hm\hn}  = \partial_{\hm}  {\mathsf V} _{\hn} - \partial_{\hn}  {\mathsf V} _{\hm} = \frac{1}{2} {\cal W}_{\hm\hn}  - \frac{1}{2} {\bf V}_{\hm\hn} \, ,\quad {\bf V}_{\hm\hn} \equiv \partial_{\hm}\Gm_{\hr\hn}^{\hr} - \partial_{\hn}\Gm_{\hr\hm}^{\hr}\, ,
 \ee
 with 
 \be \label{STGF}
 & & \Gamma_{(\hm\hn)}^{\hr} \equiv  \frac{1}{2} (\Gm_{\hm\hn}^{\hr} + \Gm_{\hn\hm}^{\hr})\, ,\qquad \Gm_{\hm\hn}^{\hr}   \equiv  \chih_{\ha}^{\;\; \hr}  \nabla_{\hm}\chi_{\hn}^{\;\;\ha} \, .
 \ee
Here the  tensor field  $\Gm_{\hm\hn}^{\hr}(\hx)$ defines a kind of {\it spacetime gauge field} with a {\it hidden gauge symmetry}. The gauge-type gravifield behaves as a {\it Goldstone-like boson field} which transmutes the local spin gauge symmetry SP(1,5) to the global Lorentz symmetry SO(1,5). 
 
The motion of the Dirac spinor is governed by various field strengths ${\cal F}_{\hm\hn}^{\ha\hb}$, $F_{\hm\hn}^i $, ${\cal G}_{\hm\hn}^{\ha}$  and ${\cal V}_{\hm\hn}$. Here ${\cal V}_{\hm\hn}$ appears as a special field strength and its effect is distinguished from other field strengths due to an imaginary factor. 

The equations of motion for the spin and scaling gauge fields as well as the charge spin gauge field are found to be
\be   \label{EMG}
& & \btd_{\hn} ( \phi \chi   \hat{\chi}^{\hm\hm'} \hat{\chi}^{\hn\hn'} {\cal F}_{\hm'\hn' }^{\;\; \ha\hb} ) =  J^{\hm\, \ha\hb}\, ,\nonumber \\
& &  \partial_{\hn} (\phi \chi   \hat{\chi}^{\hm\hm'} \hat{\chi}^{\hn\hn'}  {\cal W}_{\hm'\hn'})  =   J^{\hm}\, ,\nonumber \\
& &  D_{\hn} (\phi \chi \hat{\chi}^{\hm\hm'} \hat{\chi}^{\hn\hn'}  F^i_{\hm'\hn'} )  = J^{\hm\, i} \, ,
\ee
with the currents given by 
\be
& & J^{\hm\, \ha\hb}  = \frac{1}{2} g_s \chi \bar{\Psi}_{-} \hat{\chi}_{\ch}^{\;\; \hm} \{ \Gamma^{\ch}\;\;  \frac{1}{2} \Sigma^{\ha\hb} \} \Psi_{-}  
 + \frac{1}{2} g_s \chi \bar{\Psi}_{-} \hat{\chi}_{\ch}^{\;\; \hm} \{ \Gamma^{\ch}\;\;  \frac{1}{2} \Sigma^{\ha\hb} \} \Psi_{-} \nonumber \\
& & \qquad \quad -  \alpha_E g_s \btd_{\hn} (\chi   \hat{\chi}^{\hm\hm'} \hat{\chi}^{\hn\hn'}   \phi^2 \chi_{\hm'\hn'}^{[\ha\hb]} ) +  \frac{1}{2} \alpha_W \chi \phi^2  \hat{\chi}^{\hm \hm'} \hat{\chi}^{\hn\hn'}  \chi_{\hn}^{\;\; [\ha} {\cal G}_{\hm'\hn'}^{\hb]}  \, , \nonumber \\
& & J^{\hm} =  - 2g_w\chi \hat{\chi}^{\hm\hn} \phi d_{\hn} \phi 
- g_w   \alpha_W  \phi^2  \chi \hat{\chi}^{\hm \hm'} \hat{\chi}^{\hn\hn'} \chi_{\hn}^{\;\; \ha}  {\cal G}_{\hm'\hn' \ha}\, , \nonumber \\
& & J^{\hm\, i} = g_c \chi  \bar{\Psi}_{-} \Gamma^{\ha} \hat{\chi}_{\ha}^{\;\;\hm} \frac{1}{2}\tau^i  \Psi_{-} \, ,
\ee
where we have used the following notations
\be
& & \chi_{\hm'\hn'}^{[\ha\hb]}   =   \chi_{\hm'}^{\;\; \ha} \chi_{\hn' }^{\;\; \hb}  - \chi_{\hm' }^{\;\; \hb} \chi_{\hn' }^{\;\; \ha}  \, ; \quad 
  \chi_{\hn}^{\;\; [\ha} {\cal G}_{\hm'\hn'}^{\;\; \hb]} = \chi_{\hn}^{\;\; \ha} {\cal G}_{\hm'\hn'}^{\;\; \hb }  - \chi_{\hn}^{\;\; \hb}  {\cal G}_{\hm'\hn' }^{\;\; \ha} \, .
\ee
It can be demonstrated that the above currents are all conserved currents which satisfy the conservation laws
\be
\btd_{\hm}J^{\hm\, \ha\hb} = 0\, , \quad \partial_{\hm} J^{\hm}  =  0\, , \quad D_{\hm}J^{\hm\, i} = 0\, ,
\ee

\subsection{ Dynamics of gravifield and spacetime with totally conserved energy-momentum tensor }

The equation of motion for the gauge-type gravifield $\chi_{\hm}^{\;\; \ha}$ is found to be 
\be  \label{EMG}
& &  \nabla_{\hn} {\bf G}^{\; \hm\hn}_{\ha} = J_{\ha}^{\;\; \hm} \, , 
\ee
with the definitions for the bicovariant tensor currents and covariant derivative as follows:
\be \label{EMGC}
& & {\bf G}^{\; \hm\hn}_{\ha}  \equiv \alpha_W \, \phi^2  \chi   \hat{\chi}^{\hm \hm'} \hat{\chi}^{\hn\hn'} {\cal G}_{\hm'\hn' \ha} \, ,          \quad \nabla_{\hn} {\bf G}^{\; \hm\hn}_{\ha} =  (\partial_{\hn} - g_w W_{\hn}) {\bf G}^{\; \hm\hn}_{\ha} + g_s  {\mathcal A}_{\hn \ha}^{\;\;\; \;\hb} {\bf G}^{\; \hm\hn}_{\hb}\, , \nn \\
& & J_{\ha}^{\;\; \hm} = - \chi \hat{\chi}_{\ha}^{\;\;\hm} {\cal L} + \frac{1}{2} \chi \hat{\chi}_{\ha}^{\; \; \hr}  \hat{\chi}_{\ch}^{\;\; \hm} 
 \bar{\Psi}_{-} \Gamma^{\ch} i {\mathcal D}_{\hr} \Psi_{-} +  \chi \hat{\chi}_{\ha}^{\;\; \hn'} \hat{\chi}^{\hm \hm'} d_{\hm'}\phi  d_{\hn'} \phi   - 2 \alpha_E g_s \chi \phi^2  \hat{\chi}_{\ch}^{\;\;\hm} \hat{\chi}_{\ha}^{\;\; \hm'}    {\cal F}_{\hm'\hn'}^{\ch\hd}\hat{\chi}_{\hd}^{\;\; \hn'}   \nonumber \\
& & \quad \quad -  \chi \hat{\chi}_{\ha}^{\;\; \hr} \hat{\chi}^{\hm \hm'} \hat{\chi}^{\hn\hn'} \phi\, [\,  F^i_{\hr\hn} F^{i}_{\hm'\hn'} + {\cal F}_{\hr\hn}^{\ch\hd} {\cal F}_{\hm'\hn'\; \ch\hd} +  {\cal W}_{\hr\hn} {\cal W}_{\hm'\hn'}  - 
\alpha_W  \phi\, {\cal G}_{\hr\hn}^{\hb} {\cal G}_{\hm'\hn' \hb}    \, ] \, .
\ee

The bicovariant vector current $J_{\ha}^{\;\; \hm}$ is not a conserved current, while it is correlated to a totally conserved energy-momentum tensor ${\cal T}^{\;\, \hm}_{\hn}$ due to the translational invariance of spacetime coordinates in the action, Eq.~(\ref{Dirac6DGM}), 
\be
 \chi_{\hn}^{\;\; \ha} J_{\ha}^{\;\; \hm} =  {\cal T}^{\;\, \hm}_{\hn} \, ,  \qquad \partial_{\hm} {\cal T}^{\;\, \hm}_{\hn} = 0 \, , 
\ee
where the totally conserved energy-momentum tensor is found to be
\be \label{EMT}
{\cal T}^{\;\, \hm}_{\hn} & = & -  \eta^{\; \hm}_{\hn} \chi  {\cal L} + \frac{1}{2} \chi  \hat{\chi}_{\ha}^{\;\; \hm} 
 \bar{\Psi}_{-} \Gamma^{\ha} i {\mathcal D}_{\hn} \Psi_{-} +  \chi  \hat{\chi}^{\hm \hm'} d_{\hm'}\phi  d_{\hn} \phi   - 2 \alpha_E g_s \chi \phi^2 \hat{\chi}_{\ha}^{\;\; \hm}    {\cal F}_{\hn\hr}^{\ha\hb}\hat{\chi}_{\hb}^{\;\; \hr}   \nonumber \\
& - & \chi  \hat{\chi}^{\hm \hm'} \hat{\chi}^{\hr\hs} \phi\, [\,  F^i_{\hm'\hr} F^{i}_{\hn\hs} + {\cal F}_{\hm'\hr}^{\ha\hb} {\cal F}_{\hn\hs\; \ch\hd} +  {\cal W}_{\hm'\hr} {\cal W}_{\hn\hs}  - \alpha_W  \phi\, {\cal G}_{\hm'\hr}^{\ha} {\cal G}_{\hn\hs \ha}    \, ] \, ,
\ee
which possesses a hidden gauge symmetry.

The equation of motion for the gravifield, Eq.~(\ref{EMG}), can be expressed in connection with the totally conserved energy-momentum tensor as follows
\be \label{EMGF}
  \partial_{\hn} {\cal G}^{\; \hm\hn}_{ \hr} - {\cal G}^{\; \hm}_{\hr} ={\cal T}^{\;\, \hm}_{\hr} \, ,
\ee
with the definitions for the spacetime tensors
\be \label{GEM2}
& & {\cal G}^{\;\hm\hn}_{ \hr} \equiv \alpha_W \phi^2  \chi  \hat{\chi}^{\hm \hm'} \hat{\chi}^{\hn\hn'}  \chi_{\hr}^{\;\; \ha}\, {\cal G}_{\hm'\hn' \ha}  = - {\cal G}^{\; \hn\hm}_{ \hr} \, , \quad {\cal G}^{\; \hm}_{\hr} \equiv \Gm_{\hn\hr}^{\hs} {\cal G}^{\; \hm\hn}_{ \hs} \, .
 \ee 
Here, ${\cal G}^{\; \hm\hn}_{ \hr}$ and $ {\cal G}^{\; \hm}_{\hr}$ are regarded as the {\it gravifield tensor} and {\it gravifield tensor current} with a hidden gauge symmetry. The spacetime gauge field $\Gm_{\hn\hr}^{\hs}$ is defined in Eq.~(\ref{STGF}).

Equation~(\ref{EMGF}) provides an equation of motion for the Goldstone-like gravifield with a hidden gauge symmetry, which is an alternative to Einstein's equation of general relativity.  Note that the gauge-invariant energy-momentum tensor  ${\cal T}^{\;\, \hm}_{\hr}$ given in Eq.~(\ref{EMT}) is a totally conserved energy-momentum tensor. It contains contributions from all fields including the gravitational effect. In general, ${\cal T}_{\hn\hm}\equiv {\cal T}^{\;\, \hr}_{\hn} \eta_{\hr\hm} $ is not symmetric, $ {\cal T}_{\hn\hm} \neq {\cal T}_{\hm\hn}$, and the equation of motion, Eq.~(\ref{EMGF}), has both symmetric and antisymmetric components. The symmetric components of the equation of gravifield in Eq.~(\ref{EMGF}) lead to a generalized Einstein equation of general relativity in the six-dimensional spacetime.

In light of the energy-momentum conservation $\partial_{\hm} {\cal T}^{\;\, \hm}_{\hr}  =  \partial_{\hm} ( J^{\;\hm}_{\ha} \chi_{\hr}^{\;\; \ha} )  =0$, we obtain the following conserved current
\be \label{GFC}
& & \partial_{\hm} {\cal G}^{\; \hm}_{\hr}  =  0 \, ,
\ee
which is considered to be an alternative conservation law for the {\it gravifield tensor current}.  From Eqs.~(\ref{EMGF} )-(\ref{GFC}),  we arrive at the following equation, 
\be \label{FID}
& &  \frac{1}{2}{\mathbf R}_{\hm\hn \hr }^{\hs} \, {\cal G}^{\; \hm\hn}_{ \hs}  - \Gm_{\hm\hr}^{\; \hn} \, {\cal T}_{\hn}^{\; \hm} = 0\, ,
\ee
where ${\mathbf R}_{\hm\hn \hr }^{\hs}$ defines a field strength for the spacetime gauge field $\Gm_{\hm\hr}^{\; \hn}$,
\be
& &  {\mathbf R}_{\hm\hn \hr }^{\hs} \equiv \partial_{\hm}\Gm_{\hn \hr}^{\hs} - \partial_{\hn}\Gm_{\hm \hr}^{\hs} + \Gm_{\hm \hs'}^{\hs} \Gm_{\hn \hr}^{\hs'} - \Gm_{\hn \hs'}^{\hs} \Gm_{\hm \hr}^{\hs'}  \, .
\ee
Equation~(\ref{FID}) indicates that the vector current made by two field strengths ${\mathbf R}_{\hm\hn \hr }^{\hs}$ and ${\cal G}^{\; \hm\hn}_{ \hs}$ is identical to the vector current made through the spacetime gauge field $\Gm_{\hm\hr}^{\; \hn}$  and the totally conserved energy-momentum tensor ${\cal T}_{\hn}^{\; \hm}$.

All equations of motion are conformal scaling gauge invariant, which is attributed to the introduction of the scalar field. It is easy to read off the equation of motion for the scalar field
\be   \label{EMS}
d_{\hm}  ( \chi \hat{\chi}^{\hm\hn}  d_{\hn}\phi ) =  J \, ,
\ee
with the scalar current
\be  \label{SC}
J &= &  - \frac{1}{4}\chi \hat{\chi}^{\hm\hm'} \hat{\chi}^{\hn\hn'} [\, F^i_{\hm\hn} F^{i}_{\hm'\hn'} 
+ {\cal W}_{\hm\hn} {\cal W}_{\hm'\hn'} + {\cal F}_{\hm\hn}^{\ha\hb} {\cal F}_{\hm'\hn'\ha\hb} \, ] \nonumber \\ 
& +&  \frac{1}{2} \chi \phi [\,\alpha_W\, \hat{\chi}^{\hm\hm'}  \hat{\chi}^{\hn\hn'}  {\cal G}_{\hm\hn}^a {\cal G}_{\hm'\hn' \ha}- 4\alpha_E g_s \hat{\chi}_{\ha}^{\;\;\hm} \hat{\chi}_{\hb}^{\;\;\hn}  {\cal F}_{\hm\hn}^{\ha\hb} - 6\lambda_s \phi\, ]\, .
\ee

\section{Geometrical Symmetry Breaking Mechanism for  Mass Generation of Dirac spinor}

   It has been shown in the previous sections that a massless Dirac spinor generates both chiral symmetry and conformal scaling symmetry, which allows us to extend the usaual spinor spin symmetry SP(1,3)$\cong$SO(1,3) and Lorentz symmetry SO(1,3) in the four-dimensional Minkowski spacetime to obtain an enlarged spinor spin gauge symmetry SP(1,5)$\cong$SO(1,5)$\cong$SU$^{\ast}$(4) and global Lorentz symmetry SO(1,5) in a six-dimensional Minkowski spacetime. In other words, to yield a massive Dirac spinor, either chiral symmetry or conformal scaling symmetry has to break down.
   
   It is clear that when the spinor spin symmetry SP(1,5)$\cong$SO(1,5) is broken down to SP(1,4)$\cong$SO(1,4), the chiral symmetry is spoiled and the Dirac spinor is expected to become massive. As a demonstration, let us consider the following background structure of spacetime by choosing an appropriate expectation value of the bulk gravifield,
\be
 & & \chi_{\hm}^{\;\; \ha}(\hx)  = \langle\chi_{\hm}^{\;\; \ha}(\hx) \rangle  + \chi_{\hm}^{'\;\, \ha}(\hx) \, , \nonumber \\ 
 & & \langle\chi_{\hm}^{\;\; \ha}(\hx) \rangle = (\xi(z)\eta_{\tm}^{\;\; \ta}, \,  \zeta(z)\eta_{6}^{\;\; 6} ) \, , \quad \xi(z)\neq \zeta(z) \, ,
\ee
with $\tm=(\mu,\, 5) \, ,\, \ta = (a,\, 5)\, , \; x_6 = z$. Here $\xi(z)\neq \zeta(z)$ provides a necessary condition for the symmetry breaking, i.e., SP(1,5)$\cong$SO(1,5) is broken down to SP(1,4)$\cong$SO(1,4). The equation of motion of the Dirac spinor in such a background structure is found to be 
\be \label{DEBG}
\xi^{-1}(z)\Gamma^{\tm}i\partial_{\tm} \psi(\hx) + \zeta^{-1}(z) i\partial_{z} \psi(\hx) +  \frac{5}{2}  \zeta^{-1}(z) i\partial_z( \ln \xi(z)) \psi(\hx)  = 0 \, .
\ee
To solve the above equation,  let us consider a type of ground state solution and ignore the Kaluza-Klein modes. Namely, the Dirac field can be factorized into the following form
\be \label{FDF}
\psi(\hx) = f(z) \psi(\tx)\, , \qquad  f(z) = \rho(z)e^{i\theta(z)} \, , \qquad  x^{\tm} = (x^{\mu}\, , \, x^5)\, ,
\ee
with the Dirac field $\psi(\tx)$ defined in a five-dimensional spacetime. Suppose that $\psi(\tx)$ satisfies the following equation 
\be \label{MDE}
\Gamma^{\tm}i\partial_{\tm}\psi(\tx) = m\, \psi(\tx) \, 
\ee
with $m$ the mass of the Dirac spinor. Substituting Eqs.~(\ref{FDF})-(\ref{MDE}) into Eq.~(\ref{DEBG}), we obtain two equations,
\be
 \partial_z\theta(z) = m\, \xi^{-1}(z) \zeta(z)\, , \qquad \partial_z\ln \rho(z) = - \frac{5}{2} \partial_z \ln \xi(z) \, .
\ee
Solving the above equations, we arrive at the solutions:
\be
\theta(z) = \theta_0 + m \int^z dz'  \xi^{-1}(z') \zeta(z')\, , \qquad \rho^2(z) = \rho^2_0\, \xi^{-5}(z)\, .
\ee

In terms of such background solutions, the action for the massless Dirac spinor in the six dimensional spacetime can be reduced, by integrating over the sixth dimension, to an action with a massive Dirac spinor in a five dimensional spacetime. In order to make the resulting theory finite for an infinitely large region of the sixth dimension, i.e., $z= (-\infty\, , \, \infty) $, it requires that 
\be
\int_{-\infty}^{\infty} dz \xi^{-1}(z) \zeta(z) = \mbox{finite}\, .
\ee
A simple function satisfying the above condition can be taken as
\be
& & \zeta(z)  = \frac{1}{\sqrt{\pi}} e^{-z^2/l_c^2}\, \xi(z) \, , \qquad  \int_{-\infty}^{\infty} dz \frac{1}{\sqrt{\pi}} e^{-z^2/l_c^2} = l_c \, , \; \;\; l_c>0 \nonumber \\
& & \theta(z) = \theta_0 + m \int_{-\infty}^{z} dz' \frac{1}{\sqrt{\pi}}  e^{-z^{'2}/l_c^2} = \theta_0 + \frac{1}{2} m\, l_c ( 1 +  \mbox{erf}(z/l_c) )\, ,
\ee 
with $\mbox{erf}(z/l_c)$ the error function, where $l_c$ plays a role as a {\it characteristic length scale} of the sixth dimension. Note that the bulk gravitational field in the sixth dimension is only considered as a background field. One should in general take into account the back reaction effect by solving the gravitational equation.

\section{Conclusions and Remarks}

We have shown that a massless Dirac spinor generates new symmetries under the transformations of chirality spin and charge spin as well as conformal scaling operations. Inspired by the coherent relation between the dimensions of spacetime and the intrinsic quantum numbers of Dirac spinors, we have demonstrated with the introduction of intrinsic W-parity that the massless Dirac spinor can be treated as a Majorana-type or Weyl-type spinor in a six-dimensional spacetime that reflects the intrinsic quantum numbers of chirality spin. A generalized Dirac equation with maximal symmetry has been derived in the six-dimensional spacetime.

Based on the framework of gravitational quantum field theory with the postulate of gauge invariance and coordinate independence~\cite{YLWU}, we have built a gravitational gauge field theory in the six-dimensional spacetime by gauging the maximal symmetry of the Dirac field. The gauge-type gravifield is introduced as a bicovariant vector field defined in the six-dimensional biframe spacetime. Such a biframe spacetime is shown to be a gravifield fiber bundle $\bf{E}$. The locally flat gravifield spacetime $\bf{G}$ is regarded as a fiber and the globally flat Minkowski  spacetime $\bf{V}$ as a base spacetime. Such a gravitational gauge field theory is governed by the spinor spin gauge symmetry group SP(1,5)$\cong$SU$^{\ast}$(4) and the charge spin gauge symmetry group SU(2) in the six-dimensional Minkowski spacetime characterized by the global Poincar\'e group P(1,5)= SO(1,5)$\ltimes P^{1,5} $. The global and local conformal scaling symmetries of the theory demand the introduction of scalar and conformal scaling gauge fields. 

We have deduced a gravitational relativistic quantum equation for the massless Dirac spinor in the six-dimensional spacetime with maximal gauge and global symmetries. The equations of motion for the gauge fields have been shown to be described by the conserved currents in the presence of gravitational effects. It has been demonstrated that the dynamics of the gravifield as a Goldstone-like boson is governed by a totally conserved energy-momentum tensor. The conservation law of the energy-momentum tensor leads to an alternative equation between the field strengths and the spacetime gauge field given in the hidden gauge formalism. The symmetric part of the totally conserved energy-momentum tensor provides a generalized Einstein equation of gravity in the six-dimensional spacetime. 

We have also demonstrated a geometrical symmetry breaking mechanism for the mass generation of the Dirac spinor. When the background structure makes the spinor spin symmetry SP(1,5)$\cong$SO(1,5) break down to SP(1,4)$\cong$SO(1,4), the chiral symmetry is automatically spoiled and the Dirac spinor becomes massive. Such a mass generation mechanism is an alternative to the Higgs mechanism. It is intriguing to study a possible correlation between the Higgs boson and the gauge field component in the sixth dimension when the theory is reduced to a lower dimensional spacetime. 

In conclusion, we have shown that the maximal symmetry of the massless Dirac spinor does lead to a more general theory in the six-dimensional spacetime. Such a theory is expected to cause some new physical effects in the presence of the gravitational interaction at a high energy scale. In particular, the quantum effects of the gravitational gauge interaction and charge spin gauge interaction need to be investigated in the six-dimensional spacetime. It is also interesting to further study its implications, including the prediction of doubly electrically charged bosons, and the existence of spinor spin gauge bosons in the six dimensional spacetime. In particular, some intrinsic properties of extra dimensions and alternative symmetry breaking mechanisms need to be explored in detail.

\centerline{{\bf Acknowledgement}}

This work was supported in part by the National Science Foundation of China (NSFC) under Grants \#No. 11690022, No.11475237, No.~11121064,  and by the Strategic Priority Research Program of the Chinese Academy of Sciences, Grant No. XDB23030100 as well as  the CAS Center for Excellence in Particle Physics (CCEPP).

\clearpage
\end{CJK*}
\end{document}